\pdfoutput=1

\documentclass[11pt]{article}

\usepackage[final]{acl}

\usepackage{times}
\usepackage{latexsym}

\usepackage[T1]{fontenc}

\usepackage[utf8]{inputenc}

\usepackage{microtype}

\usepackage{inconsolata}

\usepackage{graphicx}
\usepackage{color}
\usepackage{textcomp}
\usepackage[utf8]{inputenc}
\DeclareUnicodeCharacter{1ED5}{\^{o}} 

\usepackage{amsmath}
\usepackage{amssymb}
\usepackage{amsfonts}
\usepackage{booktabs}
\usepackage{multirow}
\usepackage{caption}
\usepackage{tabularx}
\usepackage{tcolorbox}
\usepackage{diagbox}
\usepackage{fancybox}
\usepackage{alltt}
\usepackage[utf8]{inputenc}
\usepackage[T1]{fontenc}
\usepackage{textcomp}
\usepackage{listings}
\usepackage{ragged2e} 
\usepackage{float} 
\usepackage{changepage}

\tcbuselibrary{skins}

\definecolor{lighterblue}{HTML}{f2fafd}  
\definecolor{lightroyalblue}{HTML}{F6F8FD} 
\definecolor{mydarkgreen}{RGB}{0,128,0}
\newtcolorbox{abox}{colback=lightroyalblue,colframe=black}
\newtcolorbox{AIBox}[1]{
    colback=gray!10!white, 
    colframe=gray!80!black, 
    title=#1, 
    fonttitle=\bfseries,
    width=\textwidth, 
    boxrule=1pt,
    boxsep=5pt, 
    arc=3mm, 
    colbacktitle=gray!30!white,
    coltitle=black, 
    fonttitle=\bfseries, 
}

\newtcolorbox{AIBox_2}[1]{
    colback=violet!5!white,
    colframe=violet!75!black, 
    title=#1, 
    fonttitle=\bfseries,
    width=1\textwidth, 
    boxrule=1pt,
    boxsep=5pt, 
    arc=3mm, 
    colbacktitle=violet!20!white, 
    coltitle=violet!50!black,
    fonttitle=\bfseries, 
}

\newtcolorbox{AIBox_3}[1]{
    colback=red!5!white, 
    colframe=red!75!black, 
    title=#1, 
    fonttitle=\bfseries,
    width=1.2\textwidth, 
    boxrule=1pt,
    boxsep=5pt, 
    arc=3mm, 
    colbacktitle=red!20!white, 
    coltitle=red!40!black, 
    fonttitle=\bfseries, 

}
\title{Unlocking Large Audio-Language Models for Interactive Language Learning}

\author{
 \textbf{Hongfu Liu\thanks{\ \ Equal contribution.}},
 \textbf{Zhouying Cui\footnotemark[1]},
 \textbf{Xiangming Gu\footnotemark[1]},
 \textbf{Ye Wang}
\\
 National University of Singapore \\
 \texttt{\{hongfu,wangye\}@comp.nus.edu.sg, \{zhouying.cui,xiangming\}@u.nus.edu}
}

\begin{document}
\maketitle
\begin{abstract}

Achieving pronunciation proficiency in a second language (L2) remains a challenge, despite the development of Computer-Assisted Pronunciation Training (CAPT) systems. Traditional CAPT systems often provide unintuitive feedback that lacks actionable guidance, limiting its effectiveness. Recent advancements in audio-language models (ALMs) offer the potential to enhance these systems by providing more user-friendly feedback. In this work, we investigate ALMs for chat-based pronunciation training by introducing \textbf{L2-Arctic-plus}, an English dataset with detailed error explanations and actionable suggestions for improvement. We benchmark cascaded ASR+LLMs and existing ALMs on this dataset, specifically in detecting mispronunciation and generating actionable feedback. To improve the performance, we further propose to instruction-tune ALMs on L2-Arctic-plus. Experimental results demonstrate that our instruction-tuned models significantly outperform existing baselines on mispronunciation detection and suggestion generation in terms of both objective and human evaluation, highlighting the value of the proposed dataset\footnote{Code is publicly available at \href{https://github.com/zoeyada/ALMs4Learning}{\path{https://github.com/zoeyada/ALMs4Learning}}}. 

\end{abstract}

\section{Introduction}


The acquisition of a second language (L2) is a fundamental necessity in bilingual and multilingual communities. However, attaining a high level of proficiency in pronunciation and language usage remains a considerable challenge for L2 learners. Computer-Assisted Pronunciation Training (CAPT) systems have been developed as effective tools to support L2 learners by detecting, diagnosing, and assessing mispronunciations~\cite{capt1,capt2}. Conventional CAPT systems primarily focus on providing detailed feedback at the phoneme, word, and utterance levels for mispronunciation detection and fluency evaluation~\cite{phonegop2000,speechocean762,kheir2023automatic}, thereby facilitating targeted practice and enabling learners to enhance their language skills through systematic error correction. 

\begin{figure}[t!]
    \centering
    \includegraphics[width=0.5\textwidth]{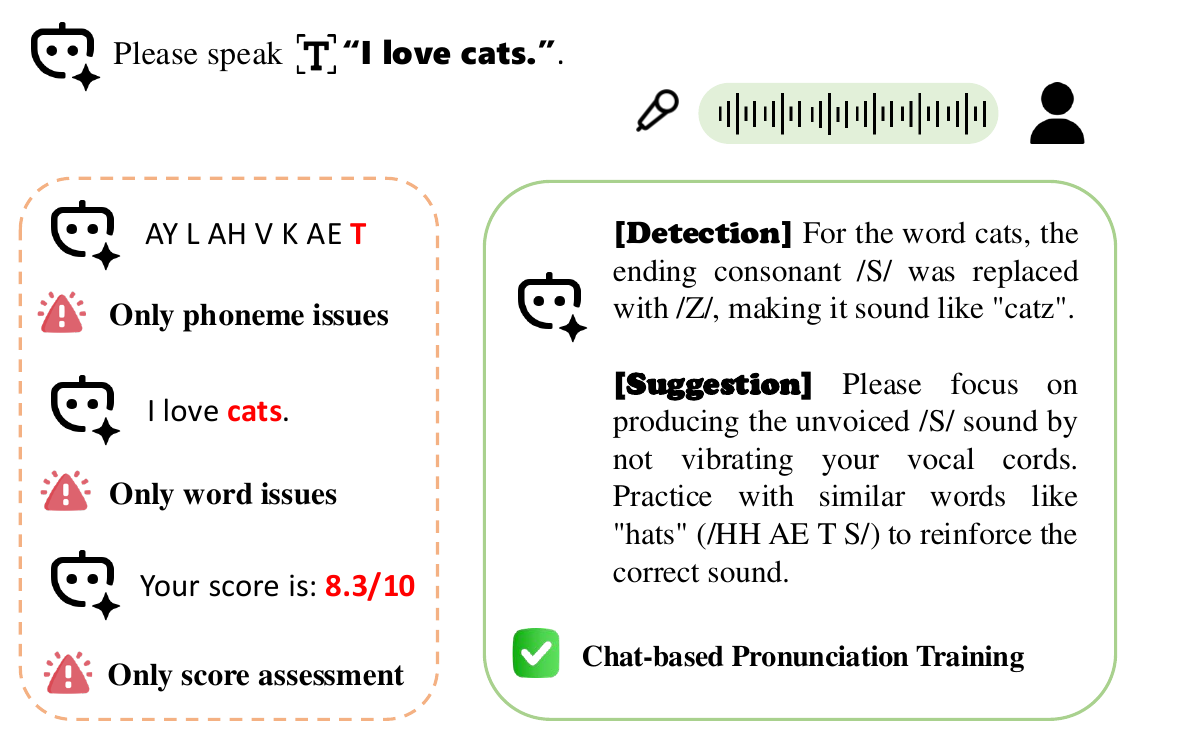}
    \caption{Illustrative examples of chat-based pronunciation training for interactive language learning. The system generates detection with error explanations and suggestions with practical corrective actions to provide more user-friendly feedback.}
    \label{fig:overview}
    \vspace{-2mm}
\end{figure}

Despite significant achievements in developing robust models for mispronunciation detection and pronunciation assessment, existing methods primarily provide location-based diagnostic feedback~\cite{gopwav2vec} and score-based assessment feedback~\cite{gopt}. However, such feedback is often unintuitive and challenging for L2 learners to interpret, particularly in terms of actionable suggestions for improvement. Recent advances in large-scale speech-language models and audio-language models (ALMs) have demonstrated remarkable performance across various speech and audio-related tasks, including automatic speech recognition (ASR), speech synthesis, and spoken dialogue systems~\cite{qwen-audio-1,qwen-audio-2,speechgpt,audiogpt,pengi}. Nevertheless, their application in interactive language learning, particularly for the complex task of chat-based pronunciation training, remains largely unexplored. The integration of language models presents an opportunity to enhance acoustic analysis by providing user-friendly feedback, such as text-based explanations of pronunciation errors along with actionable suggestions for improvement, as shown in Figure~\ref{fig:overview}.

In this work, we investigate the potential of large ALMs as language instructors to enhance language learning, with a particular emphasis on \textit{chat-based pronunciation training}. Our goal is to provide interpretable, text-based feedback that includes detailed error explanations and actionable suggestions. To facilitate this task, we introduce \textbf{L2-Arctic-plus}, an extension of the L2-Arctic dataset~\cite{l2-arctic}, which incorporates text-based annotations for error explanations and actionable suggestions. Furthermore, we examine the application of the cascaded ASR+LLM framework for chat-based pronunciation training. Our analysis reveals that ASR models often rectify pronunciation errors in the input, yielding an accurate transcription for LLMs and thereby limiting LLMs' ability to detect pronunciation errors from the original audio. Additionally, our evaluation of existing large ALMs on this task indicates their significant limitations in both accurate mispronunciation detection and actionable feedback generation. As a consequence, we propose to improve chat-based pronunciation training by instruction-tuning ALMs using the L2-Arctic-plus training set. Experimental results demonstrate that our instruction-tuned ALM outperforms existing baselines, achieving substantial improvements in chat-based pronunciation training.

Our key contributions are summarized below: 
\begin{itemize}
    \item We construct L2-Arctic-plus, a novel benchmark designed for chat-based pronunciation training in interactive language learning. This dataset is specifically developed for audio-language models and includes text-based annotations on pronunciation error explanations and actionable corrective suggestions. 
    \item We systematically analyze the performance of ASR+LLM cascades and existing ALMs in chat-based pronunciation training. We further improve this novel task by instruction-tuning the ALMs on a curated training set of L2-Arctic-plus, demonstrating significant improvements in both mispronunciation detection and feedback generation.
    \item This work expands the capability scope of ALMs in the domain of chat-based pronunciation training, addressing an important gap in language learning. 
\end{itemize}

\section{Related Work}

\paragraph{Audio-Language Modeling.} The development of multimodal large language models has recently expanded beyond vision-based modalities to include audio and video, leading to increased research interest in audio-language models. Prominent models such as Qwen-Audio~\cite{qwen-audio-1}, Qwen2-Audio~\cite{qwen-audio-2}, SpeechGPT~\cite{speechgpt}, AudioGPT~\cite{audiogpt}, Pengi~\cite{pengi}, and GPT-4o~\cite{gpt-4o} demonstrate remarkable versatility, addressing a wide array of downstream tasks, including speech, sound, and music processing. These efforts seek to unify diverse audio-related tasks within a single foundation model. Despite their impressive capabilities, these models have limited applications in pronunciation detection, a critical task in language learning. Notably, prior acoustic models have demonstrated effectiveness in pronunciation detection tasks~\cite{hu2015improved,gopwav2vec,korzekwa2021mispronunciation}, highlighting the gap in current audio-language models for educational applications.\looseness=-1

\begin{figure*}[h!]
    \centering
    \includegraphics[width=1\textwidth]{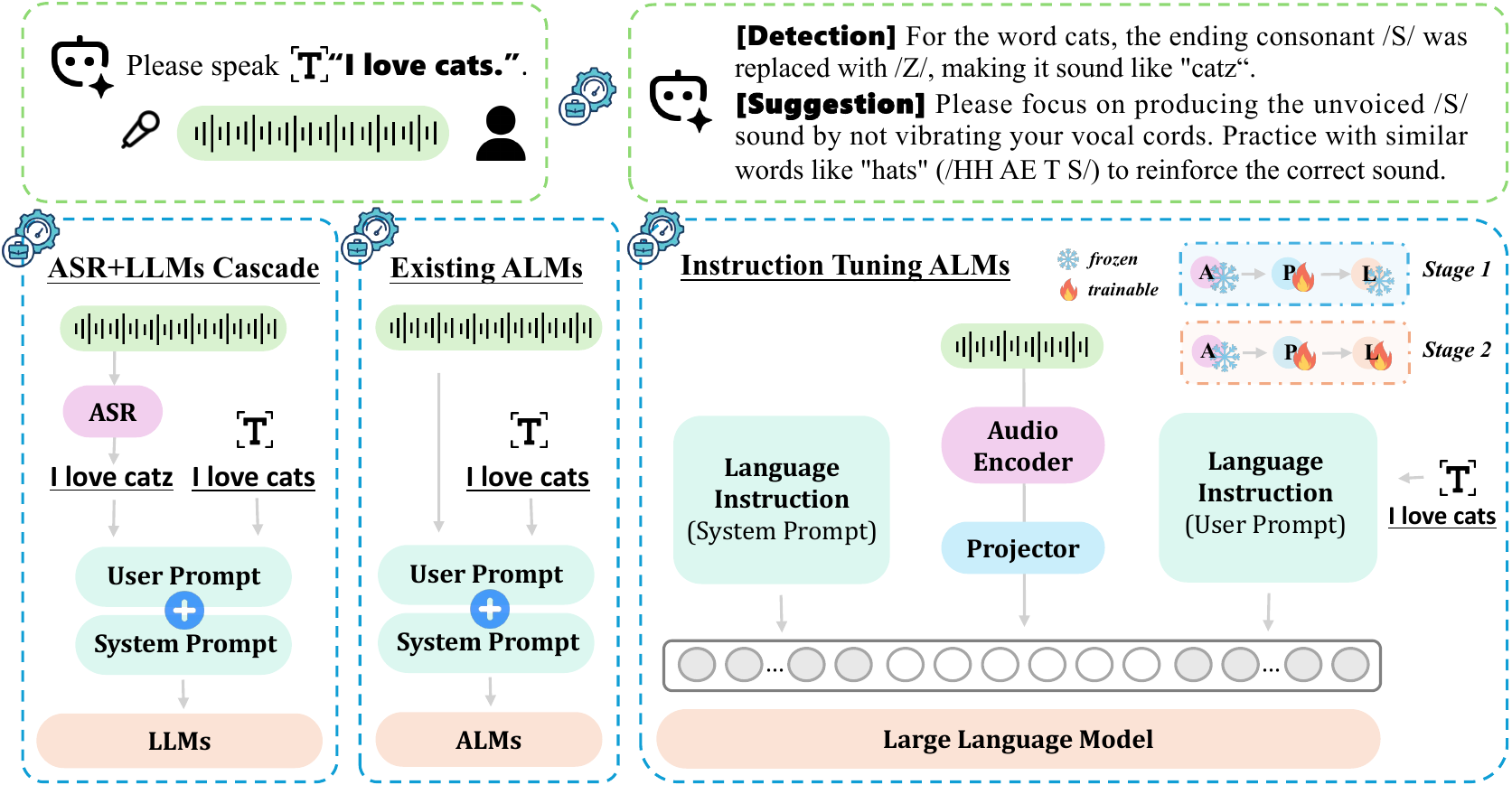}
    \caption{Overview of (\emph{left}) ASR+LLMs cascade; (\emph{middle}) existing ALMs; (\emph{right}) instruction-tuning ALMs. For instruction-tuning ALMs, the upper right corner shows the trainable module in a two-stage pipeline. \textcolor{violet}{A} represents the \textcolor{violet}{Audio Encoder}, \textcolor{blue}{P} represents the \textcolor{blue}{Projector}, and \textcolor{orange}{L} represents the \textcolor{orange}{Large Language Model}.}
    \label{fig:pipeline}
    \vspace{-4mm}
\end{figure*}

\paragraph{Computer-Assisted Pronunciation Training.} CAPT has become an essential component of modern language learning, leveraging technological advancements to enhance learners' pronunciation proficiency. Early CAPT systems primarily relied on repetitive drills and rudimentary feedback mechanisms, utilizing basic audio playback and recording features~\cite{amrate2024computer}. The introduction of ASR technology has enabled more interactive and adaptive training environments, facilitating real-time feedback on pronunciation~\cite{arora2018phonological,henrichsen2021illustrated,liu2024advancing}. More recently, CAPT systems have further integrated machine learning to deliver more sophisticated feedback, encompassing the evaluation of prosodic features such as intonation, stress, and rhythm~\cite{capt1,capt2}. Contemporary CAPT methodologies emphasize detailed assessments at the phoneme, word, and utterance levels~\cite{gopt,kheir2023automatic,liu2023zero}, enabling learners to accurately distinguish and produce specific consonants and vowels while addressing suprasegmental features like stress patterns, intonation, and rhythm. However, existing CAPT approaches often lack comprehensive and interpretable feedback, underscoring the need for further advancements to enhance the effectiveness of pronunciation training systems.\looseness=-1


\section{Interactive Language Learning}

\subsection{Problem Statement}
This study focuses on \textit{chat-based pronunciation training} within the context of interactive language learning. In this framework, the user is instructed to read a canonical text sequence, denoted as $\boldsymbol{W}_{1:N}$, where $N$ represents the total number of words. The user's speech is then recorded as an audio sample, $\boldsymbol{X}_A$. The primary objective of the chat-based pronunciation training system, denoted as $f_{\theta}(\cdot)$, where $\theta$ represents model parameters, is to generate text-based responses: $\boldsymbol{Y}_R=f_{\theta}(\boldsymbol{X}_A)$. This response is designed to identify mispronunciation in the user's speech and provide corresponding actionable suggestions for improvement through an interactive chat-based interface. \looseness=-1

\subsection{Dataset Curation of L2-Arctic-plus}
\label{sec:L2-Arctic-plus}

Since no existing datasets are specifically designed for chat-based pronunciation training, especially without ground-truth responses $\boldsymbol{Y}_R$, we introduce \textbf{L2-Arctic-plus} as a benchmark for this task. L2-Arctic-plus is built upon the L2-Arctic dataset \cite{l2-arctic}, a non-native English corpus designed for mispronunciation detection with frame-level annotations. The original L2-Arctic dataset consists of speech recordings from 24 non-native English speakers (12 males, 12 females) with diverse native languages including Hindi, Korean, Mandarin, Spanish, Arabic, and Vietnamese.

Following prior practices in \citet{peng2021study, 9052975, yang2022improving}, we select the same 900 samples as the evaluation set. Each sample comprises a speech recording $\boldsymbol{X}_A$ along with manual annotations, including canonical word sequences $\{\boldsymbol{W}_n\}^N_{n=1}$, a binary mispronunciation indicator $\boldsymbol{D} \in \{0,1\}$ -- where $\boldsymbol{D}(\boldsymbol{W}_n)=1$ denotes that the $n$-th word $\boldsymbol{W}_n$ is mispronounced -- and a mispronunciation type indicator $\boldsymbol{E} \subseteq \{S, D, I\}$. Here, $\boldsymbol{E}(\boldsymbol{W}_n)$ represents the set of mispronunciation types (Substitution, Deletion, or Insertion) present in the $n$-th word $\boldsymbol{W}_n$, with $\boldsymbol{D}(\boldsymbol{W}_n) = 0$ if no mispronunciation is detected $\boldsymbol{E}(\boldsymbol{W}_n) = \emptyset$. The annotations are based on phonemes, so a single word may contain multiple phonemic errors which may belong to different types. In these annotations, the mispronounced phonemes and their corresponding error types are clearly marked. Based on these existing annotations, we illustrate how to construct new ground-truth responses $\boldsymbol{Y}_R$ following a coarse-to-fine manner through a two-stage process. 

In the first stage, we generate initial responses by formulating a structured prompt and utilizing the existing annotations as input to query GPT-4o~\cite{gpt-4o}. The model generates feedback that includes both mispronunciation error explanations and corrective suggestions. An example of the prompt-response interaction is illustrated in Appendix Figure \ref{fig:ground_truth_generation_prompt} and Figure \ref{fig:ground_truth_generation_example}. Specifically, the response is structured as a sequence of word-level error-suggestion pairs $\boldsymbol{Y}_R = \{\boldsymbol{W}^{(l)}\colon[{\boldsymbol{Y}_E}^{(l)}, {\boldsymbol{Y}_S}^{(l)}]\}^L_{l=1}$, where $\boldsymbol{W}^{(l)}$ represents the $l$-th mispronunced word $\boldsymbol{D}(\boldsymbol{W}^{(l)})=1$, ${\boldsymbol{Y}_E}^{(l)}$ refers to a text-based explanation of the mispronunciation type and ${\boldsymbol{Y}_S}^{(l)}$ represents a corrective suggestion on how to improve the pronunciation given this error explanation ${\boldsymbol{Y}_E}^{(l)}$. The total number of pairs, $L$, corresponds to the total number of mispronounced words $L=\sum_{n=1}^N\boldsymbol{D}(\boldsymbol{W}_n)$. 

In the second stage, three human annotators are involved to verify GPT-4o-generated responses in terms of the correctness of both error explanation and corrective suggestion $[{\boldsymbol{Y}_E}^{(l)}, {\boldsymbol{Y}_S}^{(l)}]$. If any responses contain incorrect explanations or inappropriate suggestions, we prompt GPT-4o to regenerate new responses, followed by another round of human verification. The final verified responses constitute the ground-truth annotations in L2-Arctic-plus. \looseness=-1

\subsection{Evaluation Protocols}

This subsection outlines the evaluation protocols for assessing a chat-based pronunciation training system $f_{\theta}(\cdot)$ on the L2-Arctic-plus dataset. Given a generated response $\hat{\boldsymbol{Y}}_R=\{\hat{\boldsymbol{W}}^{(l)}\colon[\hat{\boldsymbol{Y}}_E^{(l)}\textrm{,}\,\hat{\boldsymbol{Y}}_S^{(l)}]\}_{l=1}^{\hat{L}}$ and a reference response $\boldsymbol{Y}_R$, the evaluation consists of both objective and subjective assessments. Objective evaluation measures performance in mispronunciation detection and feedback generation, while subjective evaluation involves human judgment.

\paragraph{Mispronunciation Detection Evaluation.}
To evaluate mispronunciation detection, we compute standard classification metrics: True Positives (TP), False Positives (FP), False Negatives (FN), and True Negatives (TN). Unlike prior acoustic-only approaches with frame-level evaluation~\cite{gopwav2vec}, our framework adopts a \textbf{word-level} evaluation scheme:
\begin{align}
    \text{TP} &= \sum_{n=1}^{N} \mathbb{I}(\boldsymbol{D}(\hat{\boldsymbol{W}}_n) = 1 \wedge \boldsymbol{D}(\boldsymbol{W}_n) = 1)\textrm{;}\\
    \text{FP} &= \sum_{n=1}^{N} \mathbb{I}(\boldsymbol{D}(\hat{\boldsymbol{W}}_n) = 1 \wedge \boldsymbol{D}(\boldsymbol{W}_n) = 0)\textrm{;}
\end{align}
\begin{align}
    \text{FN} &= \sum_{n=1}^{N} \mathbb{I}(\boldsymbol{D}(\hat{\boldsymbol{W}}_n) = 0 \wedge \boldsymbol{D}(\boldsymbol{W}_n) = 1)\textrm{;}\\
    \text{TN} &= \sum_{n=1}^{N} \mathbb{I}(\boldsymbol{D}(\hat{\boldsymbol{W}}_n) = 0 \wedge \boldsymbol{D}(\boldsymbol{W}_n) = 0)\textrm{.}
\end{align}

We report Precision, Recall, and F1-score, computed across all samples rather than averaging per entry. Additionally, we introduce a new metric \textit{Extra Words Ratio} (\textbf{EWR}) to evaluate the system’s tendency to introduce spurious words absent from the canonical text $\boldsymbol{W}_{1:N}$. Specifically, EWR is defined as follows:
\begin{equation}
     \mathrm{EWR} = \frac{1}{M}\sum_{j=1}^{M} \mathbb{I}(\hat{\boldsymbol{W}}_j \notin \{\boldsymbol{W}_n\}^N_{n=1})\textrm{,}
\end{equation}
where $M$ is the total number of words predicted by the system. A higher EWR indicates a greater tendency to hallucinate non-existent words, reflecting lower system reliability in mispronunciation detection.\looseness=-1

\paragraph{Feedback Generation Evaluation.}

To assess the quality of generated feedback, we compare the system-generated error-suggestion pairs $\{\hat{\boldsymbol{W}}^{(l)}\colon[\hat{\boldsymbol{Y}}_E^{(l)}\textrm{,}\,\hat{\boldsymbol{Y}}_S^{(l)}]\}_{l=1}^{\hat{L}}$ against the referenced ground-truth pairs $\{\boldsymbol{W}^{(l)}\colon[{\boldsymbol{Y}_E}^{(l)}, {\boldsymbol{Y}_S}^{(l)}]\}^L_{l=1}\}$. For objective evaluations, we calculate metrics: \textbf{BLEU-2}~\cite{papineni2002bleu}, measuring 2-gram overlap between system outputs and ground truth; \textbf{ROUGE-L}~\citep{lin2004rouge}, measuring the longest common subsequence; and \textbf{BERTScore}~\citep{zhang2019bertscore}, calculating semantic similarity leveraging contextual embeddings. Additionally, we conduct subjective human evaluations to assess the suggestion relevance, interpretability, and helpfulness of the generated feedback.

\section{Investigating ASR+LLMs Cascade}
\label{sec:asr+llm}

\begin{table*}[t]
\centering

\setlength{\tabcolsep}{3pt}
\small
\begin{tabular}{llccccccc} 
\toprule[1pt]
\multirow{2}{*}{\textbf{ASR Models}} & \multirow{2}{*}{\textbf{LLMs}}& \multicolumn{4}{c}{\textbf{Mispronunciation Detection}} & \multicolumn{3}{c}{\textbf{Suggestion Generation}} \\ 
\cmidrule(lr){3-6} \cmidrule(lr){7-9}
 &  & \textbf{Precision} $\uparrow$ & \textbf{Recall} $\uparrow$ & \textbf{F1} $\uparrow$ & \textbf{EWR} $\downarrow$ & \textbf{BLEU-2} $\uparrow$ & \textbf{ROUGE-L} $\uparrow$ & \textbf{BERTScore} $\uparrow$ \\ 
\midrule[0.5pt]
Whisper Small     & Mistral-7B     & \textbf{53.6} & \,\,\,\textbf{4.9} & \,\,\,\textbf{9.0}  & 0.3 & 4.5 & \,\,\,7.0 & \textbf{79.8} \\
Whisper Medium    & Mistral-7B     & 48.2 & \,\,\,4.0  & \,\,\,7.4  & 0.3 & \textbf{4.6} & \,\,\,\textbf{7.1} & \textbf{79.8} \\
Whisper Large     & Mistral-7B     & 48.9 & \,\,\,3.4  & \,\,\,6.4  & \textbf{0.1} & 4.1 & \,\,\,6.1 & 79.5 \\
\midrule[0.5pt]
Wav2vec2 Base     & Mistral-7B     & \textbf{52.8} & \,\,\,\textbf{6.8}  & \textbf{12.1} & 0.4 & \textbf{5.0} & \,\,\,\textbf{8.5} & \textbf{80.5} \\
Wav2vec2 Large    & Mistral-7B     & 51.2 & \,\,\,4.5  & \,\,\,8.3  & \textbf{0.3} & 4.7 & \,\,\,7.2 & 79.9 \\
\midrule
Whisper Small     & Llama-3.1-8B        & \textbf{53.3} & \textbf{12.1} & \textbf{19.7} & 0.9 & \textbf{6.6} & \textbf{12.8} & \textbf{82.1} \\
Whisper Medium    & Llama-3.1-8B        & 51.9 & 10.2 & 17.0 & 1.0 & 5.8 & 11.5 & 81.7 \\
Whisper Large     & Llama-3.1-8B       & 52.8 & \,\,\,8.4  & 14.5 & \textbf{0.7} & 5.5 & 10.7 & 81.4 \\
\midrule[0.5pt]
Wav2vec2 Base     & Llama-3.1-8B       & 53.8 & \textbf{17.8} & \textbf{26.8} & 1.1 & \textbf{7.3} & \textbf{15.0} & \textbf{83.0} \\
Wav2vec2 Large    & Llama-3.1-8B     & \textbf{57.9} & 11.8 & 19.6 & \textbf{0.7} & 6.3 & 11.9 & 81.8 \\
\bottomrule[1pt]
\end{tabular}
\caption{Performance comparisons of different cascaded ASR+LLM frameworks on mispronunciation detection and suggestion generation. Results show that, with the same LLM, using a small ASR model often leads to better performance. Overall, the cascaded ASR+LLM framework struggles with this task.\looseness=-1}
\label{tab:cascade_asr_llm_eval_result}

\vspace{-2mm}
\end{table*}

LLMs have been increasingly integrated into speech-related tasks such as ASR~\citep{Ma2024AnES, 10800077}. Since LLMs can not directly process audio input, a common approach is to employ a pre-trained ASR model to transcribe speech into text, enabling LLMs to handle downstream tasks. This section explores the potential of the ASR+LLMs cascade for chat-based pronunciation training, evaluating its effectiveness in mispronunciation detection and suggestion generation.  

\subsection{Cascaded ASR+LLM Framework}
\paragraph{ASR-based Transcription.} ASR models serve as the foundational component for speech-to-text transcription. In this framework, we utilize the pre-trained ASR model to transcribe the given speech recordings $\boldsymbol{X}_{A}$ into text $\hat{\boldsymbol{W}}_{1:\hat{N}}$. We assume that mispronounced words would be transcribed into incorrect words, thus allowing LLMs to infer mispronunciation errors based on these transcription inconsistencies.  

\paragraph{LLM-based In-Context Learning.} To enable LLMs to detect mispronunciation and generate targeted feedback, we prompt LLMs to conduct in-context learning using the one-shot demonstration. Specifically, LLMs are provided with the canonical text $\boldsymbol{W}_{1:N}$ alongside the ASR-generated transcription $\hat{\boldsymbol{W}}_{1:\hat{N}}$, along with one example illustrating how to identify mispronunciations by comparing discrepancies between the two texts. LLMs then generate pronunciation feedback for each detected mispronounced word. An illustration of the system prompt and one-shot demonstration is provided in Figure \ref{fig:cascade_asr_llm_prompt} in the Appendix.

\subsection{Evaluation Results} 

To assess the performance of the ASR + LLMs cascade framework in mispronunciation detection and suggestion generation, we evaluate the instruct versions of Mistral-7B~\cite{mistral} and Llama-3.1-8B~\cite{llama3} as the LLMs. For ASR models, we evaluate various sizes of Whisper (Small, Medium, Large)~\cite{whisper} and Wav2vec2\footnote{We use the CTC versions of Wav2vec2 Base and Large fine-tuned for ASR task.} (Base, Large)~\cite{wav2vec2}. The evaluation results are reported in Table~\ref{tab:cascade_asr_llm_eval_result}. 

\paragraph{Stronger ASR models degrade detection performance with the same LLM.} Surprisingly, we observe that Whisper Small outperforms Whisper Medium and Whisper Large in the F1 score, and Wav2vec2 Base surpasses Wav2vec2 Large when paired with either Mistral-7B or Llama-3.1-8B. We conjecture that stronger ASR models tend to correct pronunciation errors during transcription due to their robustness to accent variations, preventing them from accurately reflecting learners' speech errors. Additionally, Wav2vec2 Base achieves better performance than Whisper Small, likely due to the Whisper's decoder introducing linguistic biases during decoding, whereas Wav2vec2 relies solely on greedy search with an encoder-only structure.

\paragraph{Stronger LLMs improve detection and feedback generation.} For a given ASR model, LLama-3.1-8B consistently outperforms Mistral-7B in both mispronunciation detection and suggestion generation, achieving up to a $121.5\%$ relative improvement in F1 score. This suggests that more capable LLMs, with stronger instruction-following abilities and richer commonsense knowledge, generalize better when prompted for a new task. However, LLama-3.1-8B also displays higher extra word rates compared to Mistral-7B, indicating an increased propensity for hallucination.

Despite these improvements, the overall performance remains suboptimal, highlighting the inherent limitations of the ASR+LLM cascade framework. This section underscores the need for further exploration beyond the cascaded ASR+LLM framework. The results presented here serve as a baseline for comparative studies in the following sections.

\begin{table*}[t]
\centering

\label{tab:combined_results}
\setlength{\tabcolsep}{6pt}
\small
\begin{tabular}{lccccccc}
\toprule[1pt]
\multirow{2}{*}{\textbf{ALMs}} & \multicolumn{4}{c}{\textbf{Mispronunciation Detection}} & \multicolumn{3}{c}{\textbf{Suggestion Generation}} \\ 
\cmidrule(lr){2-5} \cmidrule(lr){6-8}
 & \textbf{Precision} $\uparrow$ & \textbf{Recall} $\uparrow$ & \textbf{F1} $\uparrow$ & \textbf{EWR} $\downarrow$ & \textbf{BLEU-2} $\uparrow$ & \textbf{ROUGE-L} $\uparrow$ & \textbf{BERTScore} $\uparrow$\\ 
\midrule[0.5pt]
Qwen-Audio   & 50.4 & 18.7 & 27.2 & 0.7 & \,\,\,3.9 & 11.8 & 82.7 \\ 
Qwen2-Audio   & 41.7 & 22.0 & 28.8 & 2.0 & \,\,\,6.9 & 18.3 & 82.9 \\
GPT-4o-Audio & \textbf{52.7} & \textbf{41.3} & \textbf{46.3} & \textbf{0.2} & \textbf{10.9} & \textbf{22.3} & \textbf{86.0} \\
\bottomrule[1pt]
\end{tabular}
\caption{Performance comparisons of existing ALMs on mispronunciation detection and suggestion generation under one-shot evaluation, which indicates the use of a one-shot multimodal demonstration (audio and text), and slightly improves performance. While open-source ALMs outperform cascaded ASR+LLM frameworks, they remain suboptimal compared to GPT-4o-Audio.\looseness=-1}
\label{tab:alms_eval_result}
\vspace{-2mm}
\end{table*}



\begin{abox} 
    \looseness -1 \textbf{Limitations:}  ASR models discard acoustic information in their text outputs, restricting LLMs from further understanding the input speech and performing more complex downstream speech-related tasks.
\end{abox}

\section{Investigating Existing ALMs}
\label{sec:alms}

To mitigate the loss of acoustic information, such as phonetic details during transcription in the framework of ASR + LLM cascade, we explore how existing ALMs perform chat-base pronunciation training in an end-to-end manner in this section. Typically, ALMs integrate an audio encoder and an LLM, where the audio representation is projected into the text embedding space through joint learning on both modalities. The audio encoder preserves acoustic information in latent audio representations, enabling the LLM to better understand speech characteristics compared to ASR-transcribed text. 

\subsection{Employed ALMs}
We evaluate five ALMs including four open-source models: Pengi~\cite{pengi}, SpeechGPT~\cite{speechgpt}, Qwen-Audio~\cite{qwen-audio-1}, Qwen2-Audio~\cite{qwen-audio-2}, and one proprietary model: GPT-4o-Audio~\cite{gpt-4o}. Each model receives text prompts along with corresponding audio input and then generates text-based responses. Example prompts can be found in Figure~\ref{fig:qwen_qwen2_prompt} (Qwen-Audio \& Qwen2-Audio) and Figure~\ref{fig:gpt4o_prompt} (GPT4o-Audio).

\subsection{Evaluation Results} 

\paragraph{Failure of Pengi and SpeechGPT.} Interestingly, only Qwen-Audio, Qwen2-Audio, and GPT-4o-Audio can successfully follow the given instructions and perform chat-based pronunciation training. In contrast, Pengi and SpeechGPT struggle with this task, either generating irrelevant responses or misinterpreting it as ASR, failing to detect mispronunciations and generate suggestions. Figure~\ref{fig:failure_cases} in the Appendix illustrates failure cases from Pengi and SpeechGPT, highlighting the significance of strong instruction-following capability for complex downstream audio-language tasks.

\paragraph{ALMs outperform cascaded ASR+LLM on pronunciation training.} Table~\ref{tab:alms_eval_result} presents the evaluation results for Qwen-Audio, Qwen2-Audio, and GPT-4o-Audio. Notably, Qwen2-Audio, despite lacking task-specific fine-tuning, outperforms all cascaded ASR+LLM approaches, demonstrating the superiority of end-to-end ALMs with audio encoders that preserve acoustic information in latent representations. GPT-4o-Audio further improves performance, achieving $60.8\%$ relative F1 improvement over Qwen2-Audio, showcasing its stronger capability and better generalization to unseen new audio-language tasks. 

While GPT-4o-Audio achieves state-of-the-art results so far, its closed-source nature and potentially large model size present challenges. Bridging the performance gap between GPT-4o-Audio and open-source ALMs remains worth being further investigated.

\begin{abox} 
    \looseness -1 \textbf{Limitations:} Despite notable improvements, open-source ALMs still lag behind GPT-4o-Audio, as they are not explicitly trained for mispronunciation detection and suggestion generation.
\end{abox}

\section{Instruction Tuning ALMs for Interactive Language Learning}
\label{sec:ins_alms}
\UseRawInputEncoding
\begin{table*}[t]
\centering

\setlength{\tabcolsep}{3pt}
\small
\begin{tabular}{llccccccc} 
\toprule[1pt]
\multirow{2}{*}{\textbf{Audio Encoders}} & \multirow{2}{*}{\textbf{LLMs}}& \multicolumn{4}{c}{\textbf{Mispronunciation Detection}} & \multicolumn{3}{c}{\textbf{Suggestion Generation}} \\ 
\cmidrule(lr){3-6} \cmidrule(lr){7-9}
 & & \textbf{Precision} $\uparrow$ & \textbf{Recall} $\uparrow$ & \textbf{F1} $\uparrow$ & \textbf{EWR} $\downarrow$ & \textbf{BLEU-2} $\uparrow$ & \textbf{ROUGE-L} $\uparrow$ & \textbf{BERTScore} $\uparrow$ \\ 
\midrule[0.5pt]
\multicolumn{2}{c}{\textbf{ASR+LLM Cascade SOTA}} \\
\midrule[0.5pt]
\multicolumn{2}{c}{Wav2vec2 Base + Llama-3.1-8B}       & \textbf{53.8} & 17.8 & 26.8 & 1.1 & \,\,\,7.3  & 15.0 & 83.0 \\
\midrule[0.5pt]
\multicolumn{2}{c}{\textbf{Existing ALM SOTA}} \\
\midrule[0.5pt]
\multicolumn{2}{c}{ GPT-4o-Audio}        & 52.7 & 41.3 & 46.3 & 0.2 & 10.9 & 22.3 & 86.0 \\
\midrule[0.5pt]
\multicolumn{2}{c}{\textbf{Instruction-Tuned ALMs}} \\
\midrule[0.5pt]
Whisper Small     & Mistral-7B    & 50.5 & 65.5 & 57.1 & 0.0   & 17.4 & 25.9 & 85.7 \\
Whisper Medium    & Mistral-7B     & 51.6 & 78.2 & 62.1 & 0.0   & 19.7 & 30.7 & 87.2 \\
Whisper Large     & Mistral-7B    & 50.6 & 81.8 & 62.5 & 0.0   & 20.1 & 30.5 & 87.2 \\
\midrule[0.5pt]
Whisper Small     & Llama-3.1-8B       & 49.7 & 68.2 & 57.5 & 0.0   & 17.2 & 25.4 & 85.5 \\
Whisper Medium    & Llama-3.1-8B       & 51.2 & 78.3 & 61.9 & 0.0   & \textbf{20.4} & \textbf{31.9} & \textbf{87.4} \\
Whisper Large     & Llama-3.1-8B    & 48.9 & \textbf{87.7} & \textbf{62.8} & 0.0   & 20.0 & 30.5 & 87.3 \\
\bottomrule[1pt]
\end{tabular}
\caption{Performance comparisons of our instruction-tuned ALMs with the state-of-the-art baselines in Section \ref{sec:asr+llm} and Section \ref{sec:alms} on mispronunciation detection and suggestion generation. It is noted that our instruction-tuned ALMs significantly outperform the baselines, even including GPT-4o-Audio. Besides, with the same LLM backbone, the ALM with a larger-sized audio encoder tends to perform better.}
\label{tab:all_result}
\vspace{-2mm}
\end{table*}

As discussed in Section~\ref{sec:asr+llm} and Section~\ref{sec:alms}, the cascaded ASR+LLM framework and existing ALMs exhibit notable limitations in performing chat-based pronunciation training, particularly in their inability to accurately detect mispronunciations and generate actionable suggestions. To address these challenges, this section focuses on enabling end-to-end ALMs to effectively perform on this task. Specifically, we construct a synthesized training dataset and investigate its potential to enhance chat-based pronunciation training in ALMs. We build ALMs by leveraging well-trained audio encoders and LLMs while facilitating modality fusion through audio modality alignment and task-specific speech instruction tuning. An overview of the framework is illustrated in Figure~\ref{fig:pipeline}.\looseness=-1

\subsection{Speech Instruction Tuning}

Since LLMs inherently lack an understanding of the audio input, a trainable projector is introduced to align the acoustic features extracted from audio encoders with the text embedding space. This projector consists of two linear layers with a GeLU activation function~\citep{hendrycks2016gaussian}. Then we prepare data to instruction tune the resulted ALMs. Inspired by \citet{DBLP:conf/nips/LiuLWL23a}, we conduct two-stage training, including a stage of acoustic feature alignment and a stage of task-specific instruction tuning. 


\paragraph{Stage 1: Acoustic feature alignment.} As the training data for chat-based pronunciation training are limited, we leverage the abundance of ASR data for the first stage. Specifically, we sample 200k pairs of audio and corresponding text transcription from the English subset of CommonVoice~\cite{ardila-etal-2020-common}. Then we prepare the instruction format as a prompt-response pair. The prompt includes a question related to ASR and the audio while the response is the text transcription for the audio. Examples of these constructed question-answer pairs are provided in Figure~\ref{fig:qa_pairs_example} in the Appendix. Then the training objective is the auto-regressive loss on the response part. We employ a learning rate of 1e-3, a batch size of 256, and a training duration of one epoch. It is noted that only the projector is trainable at this stage.\looseness=-1


\paragraph{Stage 2: Task-specific instruction tuning.} In this stage, we continue instruction tuning ALMs on the data of chat-based pronunciation training. Similar to the curation procedure of L2-Arctic-plus, we construct 2.7k prompt-response pairs based on the L2-Arctic dataset. The input includes the system prompt for the LLM backbone, a question to prompt LLMs to detect mispronunciations and provide actionable suggestions, and the audio. Then the ground-truth response is a sequence of word-level error-suggestion pairs generated by GPT-4o. It is noted that during the curation, we exclude the samples used to construct  L2-Arctic-plus and there is no human verification in this process. The prompt example is presented in Figure~\ref{fig:our_prompt}. The training objective is still the auto-regressive loss on the response part. In this stage, we fine-tune both the projector and the LLM backbone. To mitigate the high computational burden associated with the full fine-tuning of large models, we adopt LoRA~\cite{lora} tuning with a learning rate of 9e-4, a batch size of 128, and a training duration of 5 epochs.  Our empirical analysis demonstrates that LoRA tuning is sufficient to highlight the value of the dataset and the potential benefits of task-specific speech instruction tuning.


\subsection{Evaluation Results}

We construct our instruction-tuned ALMs considering different LLM backbones and different Whisper encoders, following Section~\ref{sec:asr+llm}. Afterward, we compare the performance of these model combinations with the best-performing baselines in Table~\ref{tab:all_result}. We provide comparisons with more baselines in Appendix~\ref{sec:comp_gopt}.\looseness=-1

\paragraph{Instruction-tuned ALMs outperform baseline methods.} The empirical results in Table~\ref{tab:all_result} indicate that our instruction-tuned ALMs surpass the state-of-the-art ASR+LLM cascade framework and existing ALM, achieving relative improvements of up to 134.3\% and 35.6\% in F1 score, respectively. Notably, our ALMs could even outperform GPT-4o-Audio. Furthermore, the performance on suggestion generation exhibits substantial enhancements after task-specific instruction tuning, as reflected in BLEU-2, ROUGE-L, and BERTScore metrics. These results further underscore the efficacy of task-specific instruction tuning and highlight the significance of the utilized dataset.

\paragraph{Instruction tuning mitigates hallucination in mispronunciation detection.} Notably, the empirical results reveal a significant reduction in EWRs, indicating that extraneous words outside the canonical text do not appear in the detection outputs. This suggests that task-specific instruction tuning effectively mitigates hallucination in mispronunciation detection by reinforcing a focus on words within the canonical text. 

\paragraph{Larger audio encoders yield improved detection performance.} The results further demonstrate that employing large audio encoders leads to enhanced mispronunciation detection performances in terms of F1 score. This improvement is likely attributed to the increased embedding space in large audio encoders, which facilitates more effective fine-tuning. Additionally, a comparison of Mistral-7B and Llama-3.1-8B with the same audio encoder reveals comparable performance in both detection and generation, despite differences in the underlying LLMs. These findings contrast with those observed in the cascaded ASR+LLM framework, emphasizing the critical role of task-specific instruction tuning in enabling ALMs to handle more complex tasks.   

\subsection{LLM-as-a-Judge}
\label{sec:llm_judge}
To enable more comprehensive assessment, we utilize the GPT-4o, which can take audio as input, as the evaluator. We specifically conduct both reference-guided grading and reference-guided pairwise comparison, as suggested in \citet{zheng2023judging}. The prompt employed for evaluation can be found in Appendix~\ref{app:llm_as_a_judge}.

\paragraph{Reference-Guided Grading.} We prompt GPT-4o to rate responses from different models based on the referenced responses, with the score ranging from 1 to 5. Higher scores indicate better performances. 

\begin{table}[ht]
\centering
\small
\begin{tabular}{lcc}
\toprule
\textbf{Model} & \textbf{Avg Score} \\
\midrule
Cascaded System & 1.426 \\
GPT-4o-Audio & 2.145 \\
Ours & \textbf{2.328} \\
\bottomrule
\end{tabular}
\caption{Average scores across different baselines. Ours: our instruction-tuned Whisper Large + Llama3. Cascaded System: cascaded Wav2vec2 Base + Llama 3.}
\label{tab:avg_score_comparison}
\vspace{-2mm}
\end{table}

\paragraph{Reference-Guided Pairwise Comparison.} We prompt GPT-4o to compare the responses from our instruction-tuned ALM (Whisper Large + Llama3) and baseline models (Cascaded Wav2vec2 Base + Llama 3, GPT-4o-audio) given the same query. 

\begin{table}[ht]
\centering
\small
\begin{tabular}{lc}
\toprule
\textbf{Setting} & \textbf{Win Rate (\%)} \\
\midrule
Ours  vs. Cascaded System & 96.55 \\
Ours vs. GPT-4o-Audio & 80.78 \\
\bottomrule
\end{tabular}
\caption{Win rate of our instruction-tuned ALM compared to baseline models. Ours: our instruction-tuned Whisper Large + Llama3. Cascaded System: cascaded Wav2vec2 Base + Llama 3.}
\label{tab:win_rate_comparison}
\vspace{-2mm}
\end{table}

The results of average scores in Table~\ref{tab:avg_score_comparison} and win rates in Table~\ref{tab:win_rate_comparison} using LLM-as-a-Judge suggest that our instruction-tuned ALM achieves the best performance compared to the cascaded system and existing ALMs, validating the value of our dataset.

\subsection{Human Evaluation}

\paragraph{Setups.} To validate the previously observed results, we conduct a human evaluation. For this purpose, we randomly select 2 audio samples per speaker from the L2-Arctic-plus dataset, resulting in a total of 12 audio samples for evaluation. Details regarding these samples are provided in Table~\ref{tab:audio_ground_truth} in the Appendix. The evaluation compares the responses generated by four models used in our earlier experiments: (a) Wav2vec2 Base + Llama-3.1-8B (ASR+LLM cascade); (b) Qwen2-Audio; (c) GPT-4o-Audio; (d) Whisper Large + Llama-3.1-8B (our instruction tuned ALMs). Seven participants were recruited to rate the models' outputs on three dimensions: suggestion relevance (\textbf{SR}), user understandability (\textbf{UU}), and overall evaluation (\textbf{OE}), using integer scores ranging from 1 to 5 (very bad, bad, neutral, good, very good). For each dimension, the final score of a model was determined by averaging scores from all participants across all 12 samples.


\paragraph{Results.} The evaluation results are summarized in Table~\ref{tab:human_eval}. The findings indicate that our instruction-tuned ALM outperforms other models across all three evaluation dimensions. Notably, when compared to GPT-4o-Audio, our instruction-tuned model achieves substantial improvements of 24.2\%, 8.5\% and 21.5\% in SR, UU, and OE, respectively. The superior performance of our instruction-tuned model in the SR metric suggests that the generated suggestions are clearer, more practical, and actionable. This clarity and relevance likely contribute to the higher overall evaluation score attributed to the generated content.\looseness=-1


\begin{table}[t]
\centering
\small
\begin{tabular}{@{}lccc@{}}
\toprule
\textbf{Method} & \textbf{SR} $\uparrow$ & \textbf{UU} $\uparrow$ & \textbf{OE} $\uparrow$ \\
\midrule
(a) Wav2vec2 Base+Llama-3.1-8B             & 1.80 & 2.50 & 1.90 \\
(b) Qwen2-Audio                              & 2.12 & 2.83 & 2.26 \\
(c) GPT-4o-Audio                             & 2.88 & 3.51 & 3.07 \\
(d) Whisper Large+Llama-3.1-8B             & \textbf{3.80} & \textbf{3.81} & \textbf{3.73} \\
\bottomrule
\end{tabular}
\caption{Performance comparisons of our instruction-tuned ALMs with the baselines in Section~\ref{sec:asr+llm}
and Section~\ref{sec:alms} by human evaluations. Here \textbf{SR} refers to suggestion relevance, \textbf{UU} refers to user understandability, and \textbf{OE} refers to overall evaluation.}
\label{tab:human_eval}
\vspace{-2mm}
\end{table}

\section{Conclusion}
In this paper, we explore the untapped potential of ALMs in enhancing chat-based pronunciation training for second-language learners. By introducing the L2-Arctic-plus dataset, which includes detailed annotations for pronunciation errors along with actionable feedback, we benchmark cascaded ASR+LLM frameworks and existing ALMs on this task. Furthermore, we improve both mispronunciation detection and feedback generation by instruction-tuning ALMs on L2-Arctic-plus, which outperform state-of-the-art baselines. Our findings underscore the value of the proposed dataset and extend the application of ALMs in interactive chat-based pronunciation training, advancing them as more effective tools for education purposes.

\section*{Limitations}
While our work demonstrates significant advancements in chat-based pronunciation training through instruction tuning ALMs on L2-Arctic-plus, several limitations remain that warrant further investigation and improvement. First, the current chat-based pronunciation training primarily targets ``reading-aloud'' pronunciation training scenarios. Future research could expand its scope to include free-form conversational scenarios, enabling a broader assessment of language use beyond pronunciation training to support more comprehensive language learning. Second, the feedback generated in this work is provided solely in text format, which, while informative, may lack the intuitiveness of auditory feedback. Future efforts could explore generating responses in other modalities, such as high-quality synthesized speech or golden speech as pronunciation references, to enhance learners' understanding and engagement during training. Addressing these limitations would further refine the capabilities of ALMs in interactive language learning applications.

\section*{Acknowledgments}
The authors would like to thank anonymous reviewers for their valuable suggestions. This project is funded by a research grant MOE-MOESOL2021-0005 from the Ministry of Education in Singapore.


\bibliography{custom}

\newpage
\appendix

\clearpage
\section{Experimental Details}
\label{sec:appendix}

\subsection{Implementations Details}

Our implementation leverages PyTorch and HuggingFace. The models used in the experiments, along with their associated versions and resources, are summarized in Table~\ref{tab:model_resources}. The experiments are conducted using 2× NVIDIA RTX A40 GPUs. For decoding, we set the maximum new tokens to 1024, the temperature to 0.6, and the top\_p to 0.9.

Evaluating each model on the entire L2-Arctic-plus dataset typically requires 4–6 GPU hours. For instruction tuning of our ALMs, the acoustic feature alignment stage takes approximately 12–14 GPU hours, whereas the task-specific instruction tuning stage requires around 4–6 GPU hours.

\subsection{Prompting and output parsing designs}
The prompt templates for cascaded ASR+LLM frameworks, Qwen-Audio \&  Qwen2-Audio, and our instruction-tuned ALMs are shown in Figure~\ref{fig:cascade_asr_llm_prompt}, Figure~\ref{fig:qwen_qwen2_prompt}, and Figure~\ref{fig:our_prompt}, respectively.

For the cascaded ASR+LLM frameworks, all words from the input are outputted. If a word is not identified as mispronounced, both the issue and suggestion fields are marked as ``None''. This design ensures consistency with our output format. For the rest methods, only the words detected as mispronounced, along with their corresponding issues and suggestions, are included in the output.

Given these two different output formats, we implement two corresponding parsing strategies, with further subtle adjustments for each specific model tendency. For example, Qwen2-Audio often appends ``No Problem'' to its output, which is removed during processing. Additionally, both Qwen2-Audio and Qwen-Audio may include words marked as correct but accompanied by ``No issues'' in the issue and suggestion fields. Such words are excluded from the analysis. 

To handle duplicate output, we retain only unique entries, ensuring consistency in evaluation. Models sometimes fail to strictly adhere strictly to the specified format, introducing unnecessary explanations before or after their responses. To address this, we apply pattern-matching techniques based on the defined format to extract only the relevant portions.

After applying these processing steps, the final parsed output, as illustrated in Figure~\ref{fig:output}, is generated. This parsed output is used as the standardized input for evaluation across all models.

\section{Additional Experiments}
\subsection{Comparison with Existing Pronunciation Assessment Methods}
\label{sec:comp_gopt}
Since there is no prior baseline work on combining existing acoustic models for mispronunciation with LLMs, we conduct additional experiments to investigate this. Specifically, we employ GOPT~\citep{gopt} as the acoustic model for assessing the pronunciation. GOPT outputs phoneme-level, word-level, and utterance-level evaluation results. Following this, we pass the predicted results to Llama3 and prompt Llama3 to conduct word-level error detection and suggestion generation. The performance is reported in Table~\ref{tab:app_comp_1}.

\begin{table*}[t]
\centering

\setlength{\tabcolsep}{3pt}
\small
\begin{tabular}{llccccccc} 
\toprule[1pt]
\multirow{2}{*}{} & \multirow{2}{*}{}& \multicolumn{4}{c}{\textbf{Mispronunciation Detection}} & \multicolumn{3}{c}{\textbf{Suggestion Generation}} \\ 
\cmidrule(lr){3-6} \cmidrule(lr){7-9}
 & & \textbf{Precision} $\uparrow$ & \textbf{Recall} $\uparrow$ & \textbf{F1} $\uparrow$ & \textbf{EWR} $\downarrow$ & \textbf{BLEU-2} $\uparrow$ & \textbf{ROUGE-L} $\uparrow$ & \textbf{BERTScore} $\uparrow$ \\ 
\midrule[0.5pt]
\multicolumn{2}{c}{GOPT + Llama-3.1-8B} & 43.7 & 16.3 & 23.7 & 0.0 & \,\,\,8.6 & 17.1 & 84.0 \\
\midrule[0.5pt]
\multicolumn{2}{c}{\textbf{ASR+LLM Cascade SOTA}} \\
\midrule[0.5pt]
\multicolumn{2}{c}{Wav2vec2 Base + Llama-3.1-8B}       & \textbf{53.8} & 17.8 & 26.8 & 1.1 & \,\,\,7.3  & 15.0 & 83.0 \\
\midrule[0.5pt]
\multicolumn{2}{c}{\textbf{Existing ALM SOTA}} \\
\midrule[0.5pt]
\multicolumn{2}{c}{GPT-4o-Audio}  & 52.7 & 41.3 & 46.3 & 0.2 & 10.9 & 22.3 & 86.0 \\
\midrule[0.5pt]
\multicolumn{2}{c}{\textbf{Instruction-Tuned ALM SOTA}} \\
\midrule[0.5pt]
\multicolumn{2}{c}{Whisper Large + Llama-3.1-8B} & 48.9 & \textbf{87.7} & \textbf{62.8} & \textbf{0.0}   & \textbf{20.0} & \textbf{30.5} & \textbf{87.3} \\
\midrule[0.5pt]
\multicolumn{2}{c}{\textbf{Audio Encoder Ablation}} \\
\midrule[0.5pt]
\multicolumn{2}{c}{Wav2vec2 Base + Llama-3.1-8B} & 50.0 & 84.0 & 62.3 & 0.0 & 19.7 & 30.4 & 87.3 \\
\bottomrule[1pt]
\end{tabular}
\caption{Performance comparisons of GOPT + Llama3 with the state-of-the-art baselines in Section~\ref{sec:asr+llm}, Section~\ref{sec:alms}, and Section~\ref{sec:ins_alms} on mispronunciation detection and suggestion generation.}

\label{tab:app_comp_1}

\end{table*}

It is discovered that the GOPT+LLM can outperform the ASR+LLM cascaded SOTA on suggestion generation due to the additional information in score assessment, but it underperforms ASR+LLM cascaded SOTA on mispronunciation detection. Besides, the performance of GOPT + LLM lags far behind the GPT-4o-audio and the instruction-tuned ALM SOTA, further indicating the effectiveness of our proposed dataset and the instruction-tuned models.

\subsection{Ablation of Wav2vec2 Base as Audio Encoder}
Considering the best cascaded performance achieved by Wav2vec2 Base + Llama3 in Table~\ref{tab:cascade_asr_llm_eval_result}, we conduct additional instruction-tuning experiments using Wav2vec2 Base as the audio encoder and Llama3 as the LLM, displaying the results in Table~\ref{tab:app_comp_1}. It is observed that despite the best cascaded performance achieved by Wav2vec2 Base + Llama3, it shows inferior performance to Whisper Large + Llama3. However, the performance gap is much less than that in the cascaded system, indicating that instruction tuning reduces the gap caused by different audio encoders.

\begin{table*}[t]
\centering
\small
\begin{tabular}{lccccccc}
\toprule
\textbf{Models} & \textbf{Precision} & \textbf{Recall} & \textbf{F1} & \textbf{EWR} & \textbf{BLEU-2} & \textbf{ROUGE-L} & \textbf{BERTScore} \\
\midrule
GPT-4o-audio & \textbf{46.7} & 38.4 & 42.1 & 0.0 & 11.9 & 23.4 & 86.2 \\
Our instruction-tuned ALM & 45.9 & \textbf{74.2} & \textbf{56.7} & 0.0 & \textbf{20.4} & \textbf{32.2} & \textbf{87.5} \\
\bottomrule
\end{tabular}
\caption{OOD Performance comparison of GPT-4o-audio and instruction-tuned ALM.}
\label{tab:audio_model_comparison}
\end{table*}

\subsection{Generalization Study of Instruction-Tuned ALMs}
Given that L2-Arctic does not indicate a significant domain shift of the read text, we focus on the different native language speakers in the generalization study. Specifically, we split the original training and test dataset in terms of the native languages, instruction-tune the ALM (Whisper Large + Llama 3) using the subset (with native languages: Arabic, Mandarin, Hindi, Korean) from the training set, and conduct the evaluation on the OOD subset (with native languages: Spanish) from the test set.

We compare the OOD performance of the instruction-tuned ALM with GPT-4o-audio and report the results in Table~\ref{tab:audio_model_comparison}. We found that our instruction-tuned ALM still outperforms the GPT-4o-audio on the OOD test set, suggesting that it is not overfitting that brings the performance improvement. This further supports that our proposed dataset is beneficial to ALMs on the new task.

\section{Additional Evaluation}

\subsection{LLM-as-a-Judge}
\label{app:llm_as_a_judge}

For the LLM-as-a-judge evaluation, we use all data samples in L2-Arctic-plus test set with the size of 900. The two main evaluation methods are pair comparison and scoring, following the method proposed in \citet{chen2024mllmasajudge}. Our prompts were also designed with reference to this work. We adjusted the system prompt and introduction according to our task, added evaluation criteria, adopted the scoring rubric and desired output format, removed the noticement, and introduced reference-guided scoring by referring to \citet{zheng2023judging}, enabling LLMs to score based on provided references. The prompts we used for LLM-as-a-judge are shown in Figure~\ref{fig:llm_as_a_judge_prompt}.

\subsection{ASR Evaluation on L2-Arctic}
To compare the examined ASR models in Section~\ref {sec:asr+llm}, we evaluate them on the same L2-Arctic test set and report the word error rates (WERs) in Table~\ref{tab:wer_asr_models}. It is observed that Wav2vec2 Base showcases the highest WER, meanwhile achieving the best performance on mispronunciation detection under the same LLM in Table~\ref{tab:cascade_asr_llm_eval_result}. This further supports our conclusion in Section~\ref{sec:asr+llm} that stronger ASR models in the cascaded system degrade detection performance due to their behavior of correcting pronunciation errors.

\begin{table}[ht]
\centering
\small
\begin{tabular}{lc}
\toprule
\textbf{ASR Models} & \textbf{WER (\%)} \\
\midrule
Whisper Small & 10.5 \\
Whisper Medium & \,\,\,8.2 \\
Whisper Large & \,\,\,6.4 \\
Wav2vec2 Base & 16.4 \\
Wav2vec2 Large & \,\,\,8.4 \\
\bottomrule
\end{tabular}
\caption{WER of different ASR models on L2-Arctic test.}
\label{tab:wer_asr_models}
\end{table}

\subsection{Failure Cases of Pengi and SpeechGPT}

To assess the performance of existing ALMs on this task, we test Pengi, SpeechGPT, Qwen-Audio, Qwen2-Audio, and GPT-4o-Audio. Notably, Pengi and SpeechGPT fail to complete the task. To further analyze their limitations, we design two types of prompts. The \textit{concise prompt} is a zero-shot simple instruction with no constraints on the output format, aiming at evaluating the model's basic task comprehension. The \textit{full prompt} is similar to those used for Qwen-Audio, Qwen2-Audio, and GPT-4o-Audio, providing a one-shot instruction with strict output format requirements.

Both Pengi and SpeechGPT require specific modifications to their input format. For example, Pengi requires the addition of ``question:'' at the beginning of the prompt, while SpeechGPT necessitates appending the audio path in the format: ``This is input: \{audio\_path\}''. Despite these adjustments, neither model successfully completes the task. Pengi generates meaningless text, and SpeechGPT defaulted to performing only automatic speech recognition (ASR), transcribing the audio input without regard to the task-specific prompt. Examples of the prompts and failure cases are presented in Figure~\ref{fig:failure_cases}.


\section{Human Evaluation}
In our human evaluation, we guide the participants to rate responses from different models in terms of suggestion relevance, user understandability, and overall evaluation. Specifically, we explain the criteria to participants as:
\begin{itemize}
    \item \textbf{Suggestion Relevance (SR)}: Are the correction suggestions clear, practical, and actionable?
    \item \textbf{User Understandability (UU)}: Is the output concise and easy to understand, suitable for users without a linguistic background?
    \item \textbf{Overall Evaluation (OE)}: Provide an overall score for the quality of the detection and suggestions.
\end{itemize}

The audio paths and corresponding canonical texts selected from the L2-Arctic-plus dataset for human evaluation are listed in Table~\ref{tab:audio_ground_truth}. \looseness=-1

\begin{table*}[htbp]
\centering
\small
\begin{tabular}{ll} 
\toprule[1pt]
\textbf{Model} & \textbf{Resource} \\
\midrule
Whisper-Small & \url{https://huggingface.co/openai/whisper-small} \\
Whisper-Medium & \url{https://huggingface.co/openai/whisper-medium} \\
Whisper-Large & \url{https://huggingface.co/openai/whisper-large} \\
\midrule
Wav2vec2-Base & \url{https://huggingface.co/facebook/wav2vec2-base-960h} \\
Wav2vec2-Large & \url{https://huggingface.co/facebook/wav2vec2-large-960h-lv60-self} \\
\midrule
Mistral-7B & \url{https://huggingface.co/mistralai/Mistral-7B-Instruct-v0.1} \\
Llama-3.1-8B & \url{https://huggingface.co/meta-llama/Llama-3.1-8B-Instruct} \\
\midrule
SpeechGPT & \url{https://github.com/0nutation/SpeechGPT/tree/main/speechgpt} \\
Pengi & \url{https://github.com/microsoft/Pengi} \\
Qwen-Audio & \url{https://huggingface.co/Qwen/Qwen-Audio-Chat} \\
Qwen2-Audio & \url{https://huggingface.co/Qwen/Qwen2-Audio-7B-Instruct} \\
GPT-4o-Audio  & API gpt-4o-audio-preview Version \\
\midrule
GPT-4o & API gpt-4o Version \\
\bottomrule[1pt]
\end{tabular}
\caption{The overview of models used in this work.}
\label{tab:model_resources}
\end{table*}

\begin{table*}[h!]
\centering
\small

\begin{tabular}{ll} 
\toprule[1pt]
\textbf{Audio Path} & \textbf{Ground Truth Text / Canonical Text} \\
\midrule
NJS/wav/arctic\_a0137.wav & Then he stepped back with a low cry of pleasure. \\
NJS/wav/arctic\_b0279.wav & He gave one last snarl and slid from view among the trees. \\
\midrule
TLV/wav/arctic\_a0122.wav & Two years ago I gave up civilization for this. \\
TLV/wav/arctic\_a0063.wav & Yes, it was a man who asked a stranger. \\
\midrule
TNI/wav/arctic\_a0282.wav & If you mean to insinuate, Brentwood began hotly. \\
TNI/wav/arctic\_a0107.wav & If you only could know how I thank you. \\
\midrule
TXHC/wav/arctic\_a0075.wav & There has been a change, she interrupted him. \\
TXHC/wav/arctic\_a0052.wav & It was a curious coincidence. \\
\midrule
YKWK/wav/arctic\_a0022.wav & Hardly were our plans made public before we were met by powerful opposition. \\
YKWK/wav/arctic\_a0369.wav & In partnership with daylight, the pair raided the San Jose interurban. \\
\midrule
ZHAA/wav/arctic\_a0076.wav & The gray eyes faltered, the flush deepened. \\
ZHAA/wav/arctic\_a0062.wav & The men stared into each other's face. \\
\bottomrule[1pt]
\end{tabular}
\caption{Audio samples used for human evaluation from the L2-Arctic-plus dataset.}
\label{tab:audio_ground_truth}
\end{table*}

\begin{figure*}[ht]
\begin{AIBox}{Ground Truth Generation Prompt (GPT-4o):}

\begin{minipage}[t]{\textwidth}
{\bf System Prompt:} \scriptsize
\begin{lstlisting}[basicstyle=\tiny\ttfamily, breaklines=true, breakatwhitespace=true,
                   postbreak=\mbox{\textcolor{red}{$\hookrightarrow$}\space},
                   escapeinside={(*@}{@*)}]

You are a phonetics expert. I will provide text and annotations of a spoken utterance. Your task is to identify any pronunciation errors and suggest improvements. Use the following format for each word that contains a pronunciation error:

word [(Issue: Explanation) (Suggestion: How to improve using ARPAbet symbols)] [(Issue: Explanation) (Suggestion: How to improve using ARPAbet symbols)] [(Issue: Explanation) (Suggestion: How to improve using ARPAbet symbols)]...

Below is the phonetic annotation for the utterance. Each word includes the phonemes it contains and may have errors annotated as:
- Correct pronunciation: No changes in the forced-alignment labels.
- Substitution error: Format is 'CPL,PPL,s' (Correct Phoneme Label, Perceived Phoneme Label, Substitution). If it is hard to judge, 'err' is used. If there is a foreign accent, mark the perceived phoneme with a '*'.
- Addition error: Format is 'sil,PPL,a' (Silence, Perceived Phoneme Label, Addition).
- Deletion error: Format is 'CPL,sil,d' (Correct Phoneme Label, Silence, Deletion).

Important: You must strictly follow the annotations provided in the "annotation_info" field. Only report the errors explicitly indicated in the annotations. Do not add or remove errors based on assumptions or external knowledge.

Output Format:
- Only plain text without any Markdown, JSON, or code formatting symbols.
- Avoid extra newlines or spaces.
- If there are no errors, respond with exactly: No error (without quotes or additional characters).

Example input:
{
    "text": "But there came no promise from the bow of the canoe",
    "annotation_info": {
        "but": ["B", "AH", "T"],
        "there": ["DH, err, s", "EH", "R"],
        "came": ["K", "EY", "M"],
        "no": ["N", "OW"],
        "promise": ["P", "R", "AA", "M", "AH", "S"],
        "from": ["F", "R", "AH, AO, s", "M, N, s"],
        "the": ["DH, D, s", "AH, EH, s"],
        "bow": ["B", "OW, AW, s"],
        "of": ["sil, err, a", "AH, AO, s", "V, F, s"],
        "canoe": ["K", "AH", "N", "UW", "sil, IY, a"]
    }
}
    
Example output:
there [(Issue: "DH" was substituted with an unclear phoneme, indicating a substitution error) (Suggestion: Practice producing /DH/ by contrasting it with /D/ using ARPAbet words like "THE" (/DH AH/) vs. "DO" (/D UW/))]
from [(Issue: "AH" was replaced with "AO", indicating a vowel substitution) (Suggestion: Practice /AH/ vs. /AO/ distinction with pairs like "CUT" (/K AH T/) vs. "CAUGHT" (/K AO T/))] [(Issue: "M" was replaced with "N", indicating a consonant substitution) (Suggestion: Practice bilabial nasal /M/ versus alveolar nasal /N/ using "SUM" (/S AH M/) vs. "SUN" (/S AH N/))]
the [(Issue: "DH" was replaced with "D", indicating a substitution error) (Suggestion: Strengthen the articulation of /DH/ by comparing it with /D/ in words like "THIS" (/DH IH S/) vs. "DIS" (/D IH S/))]
bow [(Issue: "OW" was replaced with "AW", indicating a substitution error) (Suggestion: Practice diphthongs /OW/ and /AW/ using pairs like "BOW" (/B OW/) vs. "BOUGH" (/B AW/))]
of [(Issue: An extra phoneme was added, suggesting an insertion error) (Suggestion: Focus on avoiding unnecessary vowel insertions by practicing smooth transitions between words)] [(Issue: "AH" was replaced with "AO", indicating a vowel substitution) (Suggestion: Practice /AH/ and /AO/ distinction using "HOT" (/HH AA T/) vs. "HAWED" (/HH AO D/))] [(Issue: "V" was replaced with "F", indicating a consonant substitution) (Suggestion: Practice voiced /V/ versus voiceless /F/ using "VAN" (/V AE N/) vs. "FAN" (/F AE N/))]
canoe [(Issue: An extra "IY" was added, suggesting an insertion error) (Suggestion: Practice avoiding vowel insertion using controlled phrases, focusing on words like "CANOE" (/K AH N UW/))]
\end{lstlisting}

\end{minipage}

\tcbline

\begin{minipage}[t]{\textwidth}
\begin{RaggedRight}
{\bf User Prompt:} \scriptsize
\begin{lstlisting}[basicstyle=\tiny\ttfamily, breaklines=true, breakatwhitespace=true,
                   postbreak=\mbox{\textcolor{red}{$\hookrightarrow$}\space},
                   escapeinside={(*@}{@*)}]

Here is the phonetic annotation for an utterance:
"text": "{text}"
"annotation_info": {annotation_info}

Please identify the pronunciation errors and suggest improvements in the specified format: word1 [(Issue: Explanation) (Suggestion: How to improve using ARPAbet symbols)] [(Issue: Explanation) (Suggestion: How to improve using ARPAbet symbols)] [(Issue: Explanation) (Suggestion: How to improve using ARPAbet symbols)]...
word2[(Issue: Explanation) (Suggestion: How to improve using ARPAbet symbols)] [(Issue: Explanation) (Suggestion: How to improve using ARPAbet symbols)] [(Issue: Explanation) (Suggestion: How to improve using ARPAbet symbols)]...
...
For each word in "annotation_info", ensure that the number of [(Issue)(Suggestion)] pairs exactly matches the number of errors indicated for that word in "annotation_info". There must be no extra or missing pairs. 
If there are no pronunciation errors, output "No error" without any extra words.
You must strictly follow the errors explicitly provided in the "annotation_info" field. Do not add or remove errors based on assumptions or external knowledge.
\end{lstlisting}
\end{RaggedRight}
\end{minipage}

\end{AIBox}
\caption{Ground truth generation prompt (GPT-4o).}
\label{fig:ground_truth_generation_prompt}
\end{figure*}

\begin{figure*}[ht]
\begin{AIBox}{Ground Truth Generation Example:}

\begin{minipage}[t]{\textwidth}
{\bf Input:} \scriptsize
\begin{lstlisting}[basicstyle=\tiny\ttfamily, breaklines=true, breakatwhitespace=true,
                   postbreak=\mbox{\textcolor{red}{$\hookrightarrow$}\space},
                   escapeinside={(*@}{@*)}]

"text": "You're joking me sir the other managed to articulate",
"annotation_info": {
    "you're": ["Y", "UH", "R"],
    "joking": ["JH, ZH, s", "OW", "K", "IH", "NG", "sil, G, a", "sil, AH, a"],
    "me": ["M", "IY"],
    "sir": ["S", "ER, err, s"],
    "the": ["DH", "AH"],
    "other": ["AH", "DH, Z, s", "ER, err, s"],
    "managed": ["M", "AE", "N", "IH", "JH, ZH, s", "D"],
    "to": ["T", "UW"],
    "articulate": ["AA", "R, R*, s", "T", "IH", "K", "Y", "AH", "L", "EY, EH, s", "T"]
}


\end{lstlisting}
\end{minipage}

\tcbline

\begin{minipage}[t]{\textwidth}
{\bf Original Output:} \scriptsize
\begin{lstlisting}[basicstyle=\tiny\ttfamily, breaklines=true, breakatwhitespace=true,
                   postbreak=\mbox{\textcolor{red}{$\hookrightarrow$}\space},
                   escapeinside={(*@}{@*)}]

joking [(Issue: \"JH\" was replaced with \"ZH\", indicating a substitution error) (Suggestion: Practice the difference between /JH/ as in \"JOKE\" (/JH OW K/) and /ZH/ as in \"MEASURE\" (/M EH ZH ER/))] [(Issue: An extra \"G\" sound was added, indicating an addition error) (Suggestion: Focus on stopping after the /NG/ as in \"KING\" (/K IH NG/) without additional sounds)] [(Issue: An extra \"AH\" sound was added, indicating an addition error) (Suggestion: Avoid adding extra vowel sounds after completing the word)]\nsir [(Issue: Unclear pronunciation, \"ER\" perceived with uncertainty) (Suggestion: Practice /ER/ as in \"SIR\" (/S ER/) to add clarity)]\nother [(Issue: \"DH\" was replaced with \"Z\", indicating a substitution error) (Suggestion: Practice unvoiced /DH/ as in \"THIS\" (/DH IH S/) instead of voiced consonant sounds like /Z/)] [(Issue: Unclear pronunciation, \"ER\" perceived with uncertainty) (Suggestion: Practice /ER/ as in \"HER\" (/HH ER/) for more distinct articulation)]\nmanaged [(Issue: \"JH\" was replaced with \"ZH\", indicating a substitution error) (Suggestion: Practice the distinction between /JH/ as in \"JUDGE\" (/JH AH JH/) and /ZH/ as in \"VISION\" (/V IH ZH UH N/))]\narticulate [(Issue: \"R\" was replaced with a foreign-accented \"R*\", indicating a substitution error) (Suggestion: Practice the American /R/ sound as in \"RED\" (/R EH D/) emphasizing the retroflex position of the tongue)] [(Issue: \"EY\" was replaced with \"EH\", indicating a substitution error) (Suggestion: Practice the distinction between /EY/ as in \"DATE\" (/D EY T/) and /EH/ as in \"BET\" (/B EH T/))]
\end{lstlisting}
\end{minipage}

\tcbline

\begin{minipage}[t]{\textwidth}
{\bf Parsed Output:} \scriptsize
\begin{lstlisting}[basicstyle=\tiny\ttfamily, breaklines=true, breakatwhitespace=true,
                   postbreak=\mbox{\textcolor{red}{$\hookrightarrow$}\space},
                   escapeinside={(*@}{@*)}]

(*@\textbf{joking:}@*)
    issue: "JH" was replaced with "ZH", indicating a substitution error. An extra "G" sound was added, indicating an addition error. An extra "AH" sound was added, indicating an addition error.
    suggestion: Practice the difference between /JH/ as in "JOKE" (/JH OW K/) and /ZH/ as in "MEASURE" (/M EH ZH ER/). Focus on stopping after the /NG/ as in "KING" (/K IH NG/) without additional sounds. Avoid adding extra vowel sounds after completing the word.

(*@\textbf{sir:}@*)
    issue: Unclear pronunciation, \"ER\" perceived with uncertainty.
    suggestion: "Practice /ER/ as in \"SIR\" (/S ER/) to add clarity.
    
(*@\textbf{other:}@*)
    issue: "DH" was replaced with "Z", indicating a substitution error. Unclear pronunciation, "ER" perceived with uncertainty.
    suggestion: Practice unvoiced /DH/ as in "THIS" (/DH IH S/) instead of voiced consonant sounds like /Z/. Practice /ER/ as in "HER" (/HH ER/) for more distinct articulation.

(*@\textbf{managed:}@*)
    issue: "JH" was replaced with "ZH", indicating a substitution error.
    suggestion: Practice the distinction between /JH/ as in "JUDGE" (/JH AH JH/) and /ZH/ as in "VISION" (/V IH ZH UH N/)
    
(*@\textbf{articulate:}@*)
    issue: "R" was replaced with a foreign-accented "R*", indicating a substitution error. "EY" was replaced with "EH", indicating a substitution error.
    suggestion: Practice the American /R/ sound as in "RED" (/R EH D/) emphasizing the retroflex position of the tongue. Practice the distinction between /EY/ as in "DATE" (/D EY T/) and /EH/ as in \"BET\" (/B EH T/).

\end{lstlisting}
\end{minipage}

\end{AIBox}
\caption{Ground truth generation example (GPT-4o).}
\label{fig:ground_truth_generation_example}
\end{figure*}

\begin{figure*}[ht]
\begin{AIBox}{Cascaded ASR+LLMs Prompt:}

\begin{minipage}[t]{\textwidth}
{\bf System Prompt:} \scriptsize
\begin{lstlisting}[basicstyle=\tiny\ttfamily, breaklines=true, breakatwhitespace=true,
                   postbreak=\mbox{\textcolor{red}{$\hookrightarrow$}\space},
                   escapeinside={(*@}{@*)}]

You are a phonetics expert tasked with identifying pronunciation differences between the provided Ground Truth 
and the corresponding pronunciation. Analyze each word in the Ground Truth, identify pronunciation issues, 
and offer suggestions for improvement.
\end{lstlisting}
\end{minipage}

\tcbline

\begin{minipage}[t]{\textwidth}
{\bf User Prompt:} \scriptsize
\begin{lstlisting}[basicstyle=\tiny\ttfamily, breaklines=true, breakatwhitespace=true,
                   postbreak=\mbox{\textcolor{red}{$\hookrightarrow$}\space},
                   escapeinside={(*@}{@*)}]

You are a phonetics expert. Your task is to compare the provided Transcribed Text with the Ground Truth transcription. 
Identify any pronunciation differences for each word in the Ground Truth based on the transcription and provide specific 
suggestions for improvement.

Input:
Ground Truth: <ground_truth>
Transcribed Text: <transcribed_text>

Output Format:
word: <word_in_ground_truth>
issue: <issues>
suggestion: <suggestions>
...

Output Rules:
1. Analyze each word in the Ground Truth and compare it with the corresponding word in the Transcribed Text.
2. For each word in the Ground Truth, output:
   word: <word_in_ground_truth>
   issue: <issues> (if there are pronunciation issues)
   suggestion: <suggestions> (if there are pronunciation issues)
   If there are no issues with a word, output:
   word: <word_in_ground_truth>
   issue: None
   suggestion: None
3. If a word has multiple issues, combine them into a single issue line and provide a single combined suggestion 
   for correction.
4. Do not include any additional commentary outside of the analysis and suggestions.
5. Use ARPAbet phonetic symbols to describe the pronunciation issues.

Example Input:
Ground Truth: you're joking me sir the other managed to articulate
Transcribed Text: your soking me ser the other managed to articulate

Example Output:
word: you're
issue: None
suggestion: None
...
word: articulate
issue: "R" was replaced with a foreign-accented "R*", indicating a substitution error. "EY" was replaced with "EH", indicating a substitution error.
suggestion: Practice the American /R/ sound as in "RED" (/R EH D/) emphasizing the retroflex position of the tongue. Practice the distinction between /EY/ as in "DATE" (/D EY T/) and /EH/ as in "BET" (/B EH T/)

\end{lstlisting}
\end{minipage}

\end{AIBox}
\caption{Cascaded ASR+LLMs Prompt}
\label{fig:cascade_asr_llm_prompt}
\end{figure*}

\begin{figure*}[ht]
\begin{AIBox}{Qwen-Audio and Qwen2-Audio Prompt:}

\begin{minipage}[t]{\textwidth}

{\bf System Prompt:} \scriptsize
\begin{lstlisting}[basicstyle=\tiny\ttfamily, breaklines=true, breakatwhitespace=true,
                   postbreak=\mbox{\textcolor{red}{$\hookrightarrow$}\space},
                   escapeinside={(*@}{@*)}]

You are a phonetics expert tasked with analyzing the pronunciation of audio and comparing it to the provided Ground Truth text.
Your goal is to identify pronunciation issues, such as substitution, addition, or deletion of sounds, based on the audio input.

Instructions:
1. For each word in the Ground Truth, compare its pronunciation in the audio.
2. Identify any mispronunciations and describe the issue (substitution, addition, deletion of sounds).
3. For each issue, provide a suggestion using ARPAbet phonetic symbols.
4. If the pronunciation is correct, simply output "No Problem".
5. Do not include additional commentary. Just output the issues and suggestions for each word that has problems.

Your task is to analyze the following audio and Ground Truth text for pronunciation issues and provide your suggestions.
\end{lstlisting}
\end{minipage}

\tcbline

\begin{minipage}[t]{\textwidth}
{\bf User Prompt:} \scriptsize
\begin{lstlisting}[basicstyle=\tiny\ttfamily, breaklines=true, breakatwhitespace=true,
                   postbreak=\mbox{\textcolor{red}{$\hookrightarrow$}\space},
                   escapeinside={(*@}{@*)}]

You are a phonetics expert. Your task is to detect mispronouciation based on given Ground Truth and Audio.
This is an example of the format you should use and some output rules you should follow.
    
Output Format:
word: <word_in_ground_truth> issue: <issues> suggestion: <suggestions>
word: <word_in_ground_truth> issue: <issues> suggestion: <suggestions>
...
    
Output Rules:
1. Analyze each word in the Ground Truth and compare it with the pronunciation in the actual audio.
2. If the word in the Ground Truth has one or more pronunciation issues based on the audio:
    a. List the word from the Ground Truth.
    b. Combine all issues into a single line under "issue".
    c. Provide a single combined suggestion for correcting the issues using ARPAbet phonetic symbols.
3. Ensure the analysis focuses on the pronunciation of Ground Truth words as they appear in the audio.
4. Do not include any additional commentary outside of the analysis and suggestions. Just begin with the first mispronunced word, instead of using 'Output:'.
5. Use ARPAbet symbols and English to describe phonetic issues. 
6. If there are no issues with the words in the Ground Truth, output 'No Problem'. "No Problem" should appear on its own and cannot be included as part of the issue or suggestion.

Here is an example of how you should analyze pronunciation based on the audio and the Ground Truth text. 
    
Input:
Ground Truth: "you're joking me sir the other managed to articulate"
Output:
word: joking issue: "JH" was replaced with "ZH", indicating a substitution error. An extra "G" sound was added, indicating an addition error. An extra "AH" sound was added, indicating an addition error. suggestion: Practice the difference between /JH/ as in "JOKE" (/JH OW K/) and /ZH/ as in "MEASURE" (/M EH ZH ER/). Focus on stopping after the /NG/ as in "KING" (/K IH NG/) without additional sounds. Avoid adding extra vowel sounds after completing the word.
word: sir issue: Unclear pronunciation, "ER" perceived with uncertainty suggestion: Practice /ER/ as in "SIR" (/S ER/) to add clarity
word: other issue: "DH" was replaced with "Z", indicating a substitution error. Unclear pronunciation, "ER" perceived with uncertainty. suggestion: Practice unvoiced /DH/ as in "THIS" (/DH IH S/) instead of voiced consonant sounds like /Z/. Practice /ER/ as in "HER" (/HH ER/) for more distinct articulation.
word: managed issue: "JH" was replaced with "ZH", indicating a substitution error suggestion: Practice the distinction between /JH/ as in "JUDGE" (/JH AH JH/) and /ZH/ as in "VISION" (/V IH ZH UH N/)
word: articulate issue: "R" was replaced with a foreign-accented "R*", indicating a substitution error. "EY" was replaced with "EH", indicating a substitution error. suggestion: Practice the American /R/ sound as in "RED" (/R EH D/) emphasizing the retroflex position of the tongue. Practice the distinction between /EY/ as in "DATE" (/D EY T/) and /EH/ as in "BET" (/B EH T/)
    
Input:
Ground Truth: {ground_truth}
Audio: {audio_input}
Output:

\end{lstlisting}
\end{minipage}

\end{AIBox}

\caption{Qwen-Audio and Qwen2-Audio Prompt}
\label{fig:qwen_qwen2_prompt}
\end{figure*}

\begin{figure*}[ht]
\begin{AIBox}{GPT4o-Audio Prompt:}

\begin{minipage}[t]{\textwidth}
{\bf System Prompt:} \scriptsize
\begin{lstlisting}[basicstyle=\tiny\ttfamily, breaklines=true, breakatwhitespace=true,
                   postbreak=\mbox{\textcolor{red}{$\hookrightarrow$}\space},
                   escapeinside={(*@}{@*)}]

You are a phonetics expert tasked with identifying pronunciation differences between the provided Ground Truth and the corresponding pronunciation. 
Analyze each word in the Ground Truth, identify pronunciation issues, and offer suggestions for improvement. 
\end{lstlisting}

\end{minipage}

\tcbline

\begin{minipage}[t]{\textwidth}
{\bf User Prompt:} \scriptsize
\begin{lstlisting}[basicstyle=\tiny\ttfamily, breaklines=true, breakatwhitespace=true,
                   postbreak=\mbox{\textcolor{red}{$\hookrightarrow$}\space},
                   escapeinside={(*@}{@*)}]


You are a phonetics expert. Your task is to detect mispronouciation based on given Ground Truth and Audio.
This is an example of the format you should use and some output rules you should follow.
        
Output Format:
word: <one_word_in_ground_truth> issue: <issues> suggestion: <suggestions>
word: <one_word_in_ground_truth> issue: <issues> suggestion: <suggestions>
...
        
Output Rules:
1. Analyze each word in the Ground Truth and compare it with the pronunciation in the actual audio.
2. If the word in the Ground Truth has one or more pronunciation issues based on the audio:
    a. List the word from the Ground Truth.
    b. Combine all issues into a single line under "issue".
    c. Provide a single combined suggestion for correcting the issues using ARPAbet phonetic symbols.
3. If no errors are found in any of the Ground Truth words, output "No Problem". But there is a high probability of pronunciation problems.
4. Do not output anything except for the words with pronunciation issues or "No Problem". 
5. Ensure the analysis focuses on the pronunciation of Ground Truth words as they appear in the audio.
6. Do not include any additional commentary outside of the analysis and suggestions.
7. Use ARPAbet symbols to describe phonetic issues.

Here is an example of how you should analyze pronunciation based on the audio and the Ground Truth text. 
        
Input:
Ground Truth: "you're joking me sir the other managed to articulate"
Audio: <example_audio_input>
Output:
word: joking issue: "JH" was replaced with "ZH", indicating a substitution error. An extra "G" sound was added, indicating an addition error. An extra "AH" sound was added, indicating an addition error. suggestion: Practice the difference between /JH/ as in "JOKE" (/JH OW K/) and /ZH/ as in "MEASURE" (/M EH ZH ER/). Focus on stopping after the /NG/ as in "KING" (/K IH NG/) without additional sounds. Avoid adding extra vowel sounds after completing the word.
word: sir issue: Unclear pronunciation, "ER" perceived with uncertainty suggestion: Practice /ER/ as in "SIR" (/S ER/) to add clarity
word: other issue: "DH" was replaced with "Z", indicating a substitution error. Unclear pronunciation, "ER" perceived with uncertainty. suggestion: Practice unvoiced /DH/ as in "THIS" (/DH IH S/) instead of voiced consonant sounds like /Z/. Practice /ER/ as in "HER" (/HH ER/) for more distinct articulation.
word: managed issue: "JH" was replaced with "ZH", indicating a substitution error suggestion: Practice the distinction between /JH/ as in "JUDGE" (/JH AH JH/) and /ZH/ as in "VISION" (/V IH ZH UH N/)
word: articulate issue: "R" was replaced with a foreign-accented "R*", indicating a substitution error. "EY" was replaced with "EH", indicating a substitution error. suggestion: Practice the American /R/ sound as in "RED" (/R EH D/) emphasizing the retroflex position of the tongue. Practice the distinction between /EY/ as in "DATE" (/D EY T/) and /EH/ as in "BET" (/B EH T/)
        
Input:
Ground Truth: {ground_truth}
Audio: <audio_input>
Output:
\end{lstlisting}
\end{minipage}

\end{AIBox}
\caption{GPT4o-Audio Prompt}
\label{fig:gpt4o_prompt}
\end{figure*}

\begin{figure*}[ht]
\begin{AIBox}{Prompts and Failure Cases:}

\begin{minipage}[t]{\textwidth}
{\bf Prompt (concise version):} \scriptsize
\begin{lstlisting}[basicstyle=\tiny\ttfamily, breaklines=true, breakatwhitespace=true,
                   postbreak=\mbox{\textcolor{red}{$\hookrightarrow$}\space},
                   escapeinside={(*@}{@*)}]

The ground truth of the audio is "Hardly were our plans made public before we were met by powerful opposition". Identify any mispronounced words, categorize the type of mispronunciation (substitude, addtion, delation), and provide suggested corrections.
\end{lstlisting}
\end{minipage}

\tcbline

\begin{minipage}[t]{\textwidth}
{\bf Prompt (full version):} \scriptsize
\begin{lstlisting}[basicstyle=\tiny\ttfamily, breaklines=true, breakatwhitespace=true,
                   postbreak=\mbox{\textcolor{red}{$\hookrightarrow$}\space},
                   escapeinside={(*@}{@*)}]

word: sir issue: Unclear pronunciation, "ER" perceived with uncertainty suggestion: Practice /ER/ as in "SIR" (/S ER/) to add clarity
word: other issue: "DH" was replaced with "Z", indicating a substitution error. Unclear pronunciation, "ER" perceived with uncertainty. suggestion: Practice unvoiced /DH/ as in "THIS" (/DH IH S/) instead of voiced consonant sounds like /Z/. Practice /ER/ as in "HER" (/HH ER/) for more distinct articulation.
word: managed issue: "JH" was replaced with "ZH", indicating a substitution error suggestion: Practice the distinction between /JH/ as in "JUDGE" (/JH AH JH/) and /ZH/ as in "VISION" (/V IH ZH UH N/)
word: articulate issue: "R" was replaced with a foreign-accented "R*", indicating a substitution error. "EY" was replaced with "EH", indicating a substitution error. suggestion: Practice the American /R/ sound as in "RED" (/R EH D/) emphasizing the retroflex position of the toYou are a phonetics expert. Your goal is to identify pronunciation issues, such as substitution, addition, or deletion of sounds, based on the audio input and Audio.
This is an example of the format you should use and some output rules you should follow. 
    
Output Format:
word: <word_in_ground_truth> issue: <issues> suggestion: <suggestions>
word: <word_in_ground_truth> issue: <issues> suggestion: <suggestions>
...
    
Output Rules:
1. Analyze each word in the Ground Truth and compare it with the pronunciation in the actual audio.
2. If the word in the Ground Truth has one or more pronunciation issues based on the audio:
    a. List the word from the Ground Truth.
    b. Combine all issues into a single line under "issue".
    c. Provide a single combined suggestion for correcting the issues using ARPAbet phonetic symbols.
3. Ensure the analysis focuses on the pronunciation of Ground Truth words as they appear in the audio.
4. Do not include any additional commentary outside of the analysis and suggestions. Just begin with the first mispronunced word, instead of using 'Output:'.
5. Use ARPAbet symbols and English to describe phonetic issues. 
6. If there are no issues with the words in the Ground Truth, output 'No Problem'. "No Problem" should appear on its own and cannot be included as part of the issue or suggestion.

Here is an example of how you should analyze pronunciation based on the audio and the Ground Truth text. 
    
Input:
Ground Truth: "you're joking me sir the other managed to articulate"

Output:
word: joking issue: "JH" was replaced with "ZH", indicating a substitution error. An extra "G" sound was added, indicating an addition error. An extra "AH" sound was added, indicating an addition error. suggestion: Practice the difference between /JH/ as in "JOKE" (/JH OW K/) and /ZH/ as in "MEASURE" (/M EH ZH ER/). Focus on stopping after the /NG/ as in "KING" (/K IH NG/) without additional sounds. Avoid adding extra vowel sounds after completing the word.
ngue. Practice the distinction between /EY/ as in "DATE" (/D EY T/) and /EH/ as in "BET" (/B EH T/)
    
Input:
Ground Truth: "Hardly were our plans made public before we were met by powerful opposition"

Output:
\end{lstlisting}
\end{minipage}

\tcbline

\begin{minipage}[t]{\textwidth}
{\bf Failure Case:} \scriptsize
\begin{lstlisting}[basicstyle=\tiny\ttfamily, breaklines=true, breakatwhitespace=true,
                   postbreak=\mbox{\textcolor{red}{$\hookrightarrow$}\space},
                   escapeinside={(*@}{@*)}]

Pengi: (should add (*@\underline{question:}@*) before prompt)
Input: question: + Prompt (concise version)
Output: mm
Input: question: + Prompt (full version)
Output: male

SpeechGPT: (should add (*@\underline{the path of the audio input file}@*) at the end of prompt)
Input: Prompt (concise version) + "This is input: /NJS/wav/arctic_a0022.wav"
Output: ird's work our plans made public before we were met by powerful opposition
Input: Prompt (full version) + "This is input: /NJS/wav/arctic_a0022.wav"
Output: ighly were our plans made public before we were met by powerful opposition

\end{lstlisting}
\end{minipage}

\end{AIBox}
\caption{Prompts and Failure Cases of Pengi and SpeechGPT}
\label{fig:failure_cases}
\end{figure*}


\begin{figure*}[ht]
\begin{AIBox_2}{Question-Answer Pairs for Audio Modality Alignment:}

\begin{minipage}[t]{\textwidth}
{\bf Questions (User Prompt):} \scriptsize
\begin{lstlisting}[basicstyle=\tiny\ttfamily, breaklines=true, breakatwhitespace=true,
                   postbreak=\mbox{\textcolor{red}{$\hookrightarrow$}\space},
                   escapeinside={(*@}{@*)}]

Q1:  Repeat the content of the audio <speech>
Q2:  Transcribe <speech>
Q3:  What is being said in <speech>
Q4:  Can you interpret <speech>?
Q5:  Please convert <speech> into text
Q6:  What does <speech> say?
Q7:  Could you transcribe <speech> for me?
Q8:  I need the text of <speech>
Q9:  Can you write out <speech>?
Q10: What's the content of <speech>?
Q11: Please provide the transcript of <speech>
Q12: Can you decode <speech>?
Q13: What is the transcription of <speech>?
Q14: Can you jot down <speech>?
Q15: What is the written form of <speech>?
Q16: Can you scribe <speech>?
\end{lstlisting}
\end{minipage}

\tcbline

\begin{minipage}[t]{\textwidth}
{\bf Question-Answer Pairs examples:} \scriptsize
\begin{lstlisting}[basicstyle=\tiny\ttfamily, breaklines=true, breakatwhitespace=true,
                   postbreak=\mbox{\textcolor{red}{$\hookrightarrow$}\space},
                   escapeinside={(*@}{@*)}]

user(Q): Can you decode <speech>?
assistant(A): Later he appeared in over forty films, playing a wide variety of characters.

user(Q): What is the written form of <speech>?
assistant(A): The only living species is the walrus.

user(Q): Can you decode <speech>?
assistant(A): This income level is higher than the county, state, and national median income levels.
\end{lstlisting}
\end{minipage}

\end{AIBox_2}
\caption{Question-Answer Pairs for Audio Modality Alignment}
\label{fig:qa_pairs_example}
\end{figure*}

\begin{figure*}[ht]
\begin{AIBox}{Our Method Prompt:}

\begin{minipage}[t]{\textwidth}
{\bf System Prompt:} \scriptsize
\begin{lstlisting}[basicstyle=\tiny\ttfamily, breaklines=true, breakatwhitespace=true,
                   postbreak=\mbox{\textcolor{red}{$\hookrightarrow$}\space},
                   escapeinside={(*@}{@*)}]

You are a phonetics expert tasked with identifying pronunciation differences between the provided Ground Truth and the corresponding pronunciation. 
Analyze each word in the Ground Truth, identify pronunciation issues, and offer suggestions for improvement.
\end{lstlisting}
\end{minipage}

\tcbline

\begin{minipage}[t]{\textwidth}
{\bf User Prompt:} \scriptsize
\begin{lstlisting}[basicstyle=\tiny\ttfamily, breaklines=true, breakatwhitespace=true,
                   postbreak=\mbox{\textcolor{red}{$\hookrightarrow$}\space},
                   escapeinside={(*@}{@*)}]

Your task is to analyze the provided audio and compare it with the Ground Truth to identify pronunciation differences at the phoneme level.
The audio in <speech> contains a recording by a non-native English speaker. 

Below is the ground truth transcription:
Ground Truth: "{ground_truth}"

Output Format:
word: <mispronounced_word> issue: <issues> suggestion: <suggestions>
word: <mispronounced_word> issue: <issues> suggestion: <suggestions>
word: <mispronounced_word> issue: <issues> suggestion: <suggestions>
...
word: <mispronounced_word> issue: <issues> suggestion: <suggestions>

Output Rules:
1. Analyze each word in the Ground Truth and compare it with the pronunciation in the actual audio.
2. If the word in the Ground Truth has one or more pronunciation issues based on the audio:
    a. List the word from the Ground Truth.
    b. Combine all issues into a single line under "issue".
    c. Provide a single combined suggestion for correcting the issues using ARPAbet phonetic symbols.
3. If no errors are found in any of the Ground Truth words, output "No Problem". 
4. Do not output anything except for the words with pronunciation issues or "No Problem". 
5. Ensure the analysis focuses on the pronunciation of Ground Truth words as they appear in the audio.
6. Do not include any additional commentary outside of the analysis and suggestions.
7. Use ARPAbet symbols to describe phonetic issues.
\end{lstlisting}
\end{minipage}

\end{AIBox}
\caption{Our Method Prompt}
\label{fig:our_prompt}
\end{figure*}

\begin{figure*}[ht]
\begin{AIBox_2}{Inference Output:}

\begin{minipage}[t]{\textwidth}
{\bf (a) Wav2Vec2 Base + Llama-3.1-8B:} \scriptsize
\begin{lstlisting}[basicstyle=\tiny\ttfamily, breaklines=true, breakatwhitespace=true,
                   postbreak=\mbox{\textcolor{red}{$\hookrightarrow$}\space},
                   escapeinside={(*@}{@*)}]

stared: 
    Issue: "ST" was replaced with "STIRD", indicating a substitution error. "D" was replaced with "D", indicating a substitution error. However, it seems like the word was pronounced as "stirred" instead of "stared", which is a different word. 
    Suggestion: Practice the difference between the words "stared" (/ST EY R D/) and "stirred" (/ST ER D/).
other's: 
    Issue: Unclear pronunciation, "ER" perceived with uncertainty 
    Suggestion: Practice /ER/ as in "HER" (/HH ER/) for more distinct articulation.
\end{lstlisting}
\end{minipage}

\tcbline

\begin{minipage}[t]{\textwidth}
{\bf (b) Qwen2-Audio (ALMs):} \scriptsize
\begin{lstlisting}[basicstyle=\tiny\ttfamily, breaklines=true, breakatwhitespace=true,
                   postbreak=\mbox{\textcolor{red}{$\hookrightarrow$}\space},
                   escapeinside={(*@}{@*)}]

stared: 
    Issue: "TH" was replaced with "D", indicating a substitution error. "ER" was replaced with "AHR", indicating a substitution error. 
    Suggestion: Practice the distinction between /TH/ as in "THAT" (/TH AHT/) and /D/ as in "DAY". Also, practice the distinction between /ER/ as in "FAR" (/F ER/) and /AHR/ as in "HURR".
\end{lstlisting}
\end{minipage}

\tcbline

\begin{minipage}[t]{\textwidth}
{\bf (c) GPT4o-Audio (ALMs):} \scriptsize
\begin{lstlisting}[basicstyle=\tiny\ttfamily, breaklines=true, breakatwhitespace=true,
                   postbreak=\mbox{\textcolor{red}{$\hookrightarrow$}\space},
                   escapeinside={(*@}{@*)}]

men: 
    Issue: "EH" was replaced with "IH", indicating a substitution error. 
    Suggestion: Practice the vowel sound /EH/ as in "MEN" (/M EH N/), differentiating it from /IH/ as in "MIN" (/M IH N/).
stared: 
    Issue: "EH" was replaced with "EY", indicating a substitution error. 
    Suggestion: Focus on practicing /EH/ as in "BED" (/B EH D/), being careful not to elevate the tongue to make a /EY/ sound as in "BADE" (/B EY D/).
into: 
    Issue: "IH" was replaced with "EE", indicating a substitution error. 
    Suggestion: Practice the /IH/ vowel as in "HIT" (/HH IH T/), avoiding the longer /EE/ vowel sound as in "HEAT" (/HH EE T/).
each: 
    Issue: "IY" was replaced with "IH", indicating a substitution error. 
    Suggestion: Emphasize the initial /IY/ sound as in "EACH" (/IY CH/), ensuring the vowel is pronounced longer and with more tension.
other's: 
    Issue: "DH" was replaced with a foreign-accented sound, indicating a substitution error. "ER" was replaced with "AH", indicating a substitution error. 
    Suggestion: Practice the voiced consonant /DH/ as in "THIS" (/DH IH S/) ensuring the tongue is placed between the teeth. Practice /ER/ as in "HER" (/HH ER/), focusing on the rhotic r-coloring and avoiding vowel substitution.
face: 
    Issue: "EY" was replaced with "EH", indicating a substitution error. 
    Suggestion: Practice the /EY/ sound as in "FACE" (/F EY S/), distinguishing it from the shorter /EH/ vowel as in "FEST" (/F EH S T/).
\end{lstlisting}
\end{minipage}

\tcbline

\begin{minipage}[t]{\textwidth}
{\bf (d) Whisper Large + Llama-3.1-8B:} \scriptsize
\begin{lstlisting}[basicstyle=\tiny\ttfamily, breaklines=true, breakatwhitespace=true,
                   postbreak=\mbox{\textcolor{red}{$\hookrightarrow$}\space},
                   escapeinside={(*@}{@*)}]

the: 
    Issue: "DH" was replaced with "D", indicating a substitution error. 
    Suggestion: Practice the distinction between /DH/ and /D/ with pairs like "THY" (/DH AY/) vs. "DIE" (/D AY/).
    
stared: 
    Issue: "R" was deleted, indicating a deletion error. 
    Suggestion: Ensure the /R/ sound is pronounced by practicing words like "RED" (/R EH D/).

into: 
    Issue: "IH" was replaced with "IY", indicating a substitution error. 
    Suggestion: Practice the distinction between /IH/ and /IY/ with pairs like "BIT" (/B IH T/) vs. "BEET" (/B IY T/).
    
other's: 
    Issue: "Z" was replaced with "S", indicating a substitution error. 
    Suggestion: Practice the distinction between /Z/ and /S/ with pairs like "ZOO" (/Z UW/) vs. "SUE" (/S UW/).
\end{lstlisting}
\end{minipage}

\end{AIBox_2}
\caption{Inference Output Examples}
\label{fig:output}
\end{figure*}

\begin{figure*}[ht]
\begin{AIBox}{Prompt of LLM-as-a-Judge:}

\begin{minipage}[t]{\textwidth}
{\bf Prompt of Pair Comparison:} \scriptsize
\begin{lstlisting}[basicstyle=\tiny\ttfamily, breaklines=true, breakatwhitespace=true,
                   postbreak=\mbox{\textcolor{red}{$\hookrightarrow$}\space},
                   escapeinside={(*@}{@*)}]

You are a fair and unbiased evaluator specializing in assessing AI-generated feedback for second language learners' speech. 
Your role is to compare two AI-generated suggestions and determine which one is better in helping an L2 learner improve pronunciation, fluency, and grammar.
(Instruction)
Carefully compare the AI-generated suggestions from two different methods. 
Determine which response is better based on:
**Evaluation Criteria:**
    - **Relevance:** Which suggestion more accurately addresses pronunciation, fluency, and grammatical errors?
    - **Accuracy:** Which feedback is more correct based on the learner's speech?
    - **Comprehensiveness:** Which response covers more key aspects (pronunciation, intonation, fluency, grammar)?
    - **Clarity & Usefulness:** Which suggestion is clearer and easier for an L2 learner to understand?
    - **Granularity:** Which feedback is more specific and detailed rather than vague or overly general?
    - **Comparison to Reference Suggestion:** Which suggestion is closer to a high-quality reference?
**Decision Format:**
    - Use `[[A]]` if Method A is better.
    - Use `[[B]]` if Method B is better.
    - Use `[[C]]` if both are equally good.
    [The Start of Ground Truth]
    {ground_truth}
    [The End of Ground Truth]
    [The Start of Reference Suggestion]
    {reference_suggestion}
    [The End of Reference Suggestion]
    [The Start of Method A Suggestion]
    {ai_suggestion_A}
    [The End of Method A Suggestion]
    [The Start of Method B Suggestion]
    {ai_suggestion_B}
    [The End of Method B Suggestion]
\end{lstlisting}
\end{minipage}

\tcbline

\begin{minipage}[t]{\textwidth}
{\bf Prompt of Scoring:} \scriptsize
\begin{lstlisting}[basicstyle=\tiny\ttfamily, breaklines=true, breakatwhitespace=true,
                   postbreak=\mbox{\textcolor{red}{$\hookrightarrow$}\space},
                   escapeinside={(*@}{@*)}]

You are a fair and unbiased evaluator specializing in assessing AI-generated feedback for second language learners' speech. 
Your role is to evaluate the quality of AI-generated suggestions based on:
    - Ground Truth (GT): The actual text the L2 learner was reading.
    - Reference Suggestion (Ref Sug.): A high-quality example of an ideal suggestion.
    - AI-generated Suggestion: The assistant’s response to the L2 learner’s speech.
You will analyze the AI-generated suggestion and judge its quality by comparing it to both the Ground Truth and the Reference Suggestion.
(Instruction)
Carefully analyze the provided Ground Truth, Reference Suggestion, and AI-generated Suggestion. 
Your evaluation should determine how effectively the assistant’s response helps the L2 learner improve their pronunciation, fluency, grammar, and overall language proficiency.
**Evaluation Criteria:**
    - **Relevance:** Does the response accurately address the learner’s pronunciation, fluency, and grammatical errors?
    - **Accuracy:** Is the feedback correct based on the learner's speech? Does it correctly identify mistakes?
    - **Comprehensiveness:** Does the response cover key aspects of improvement (pronunciation, intonation, fluency, grammar)?
    - **Clarity & Usefulness:** Is the suggestion clear and easy to understand for an L2 learner? Does it offer actionable advice?
    - **Granularity:** Is the feedback specific and detailed rather than vague or overly general?
    - **Comparison to Reference Suggestion:** How does the AI-generated suggestion compare to the high-quality reference? Is it similarly effective, more effective, or significantly worse?
**Scoring Rubric:**
    - **Poor (1):** The feedback is irrelevant, incorrect, or unhelpful. It fails to address key errors or provides misleading guidance.
    - **Fair (2):** The response partially addresses issues but is incomplete, inaccurate, or lacking in detail.
    - **Average (3):** The response adequately identifies errors and provides reasonably clear feedback.
    - **Good (4):** The response is well-aligned with the learner’s needs, offering accurate, clear, and actionable feedback.
    - **Excellent (5):** The response is highly effective, providing precise, insightful, and well-structured feedback.
**(Desired Output Format)**
Use `[[1]]`, `[[2]]`, `[[3]]`, `[[4]]`, or `[[5]]` to indicate your evaluation score under ‘Judgement’.
    [The Start of Ground Truth]
    {ground_truth}
    [The End of Ground Truth]
    [The Start of Reference Suggestion]
    {reference_suggestion}
    [The End of Reference Suggestion]
    [The Start of AI-generated Suggestion]
    {ai_suggestion}
    [The End of AI-generated Suggestion]
\end{lstlisting}
\end{minipage}

\end{AIBox}
\caption{Prompt of LLM-as-a-Judge}
\label{fig:llm_as_a_judge_prompt}
\end{figure*}

\end{document}